\documentclass[fleqn,usenatbib]{mnras}

\usepackage{newtxtext,newtxmath}

\usepackage[T1]{fontenc}

\DeclareRobustCommand{\VAN}[3]{#2}
\let\VANthebibliography\thebibliography
\def\thebibliography{\DeclareRobustCommand{\VAN}[3]{##3}\VANthebibliography}

\usepackage{graphicx}	
\usepackage{amsmath}	
\usepackage{subcaption}
\usepackage{float}
\usepackage{tablefootnote}

\newcommand{\teq}{T_{\rm eq}}
\def\K{\; {\rm K}}

\def \tdrag {\tau_\mathrm{drag}}

\title[General circulation models of ultra-hot Jupiters]{Modeling the day-night temperature variations of ultra-hot Jupiters: confronting non-grey general circulation models and observations}

\author[Tan et al.]{Xianyu Tan$^{1,2,3}$\thanks{E-mail: \url{xianyut@sjtu.edu.cn}}, Thaddeus D. Komacek$^{4}$, Natasha E. Batalha$^{5}$, Drake Deming$^{4}$, Roxana Lupu$^{6}$,  \newauthor Vivien Parmentier$^{7}$, and Raymond T. Pierrehumbert$^{3}$
\\
$^{1}$Tsung-Dao Lee Institute, Shanghai Jiao Tong University, 520 Shengrong Road, Shanghai, People's Republic of China\\
$^{2}$School of Physics and Astronomy, Shanghai Jiao Tong University, 800 Dongchuan Road, Shanghai, People's Republic of China\\
$^{3}$Atmospheric Oceanic  and Planetary Physics, Department of Physics, University of Oxford, OX1 3PU, UK\\
$^{4}$Department of Astronomy, University of Maryland, College Park, MD 20742, USA\\
$^{5}$NASA Ames Research Center, Moffett Field, CA 94035, USA\\
$^{6}$Eureka Scientific Inc, Oakland, CA 94602, USA \\
$^{7}$Université Côte d’Azur, Observatoire de la Côte d’Azur, CNRS, Laboratoire Lagrange, Nice, France
}

\date{Accepted XXX. Received YYY; in original form ZZZ}

\pubyear{2022}

\begin{document}
\label{firstpage}
\pagerange{\pageref{firstpage}--\pageref{lastpage}}
\maketitle

\begin{abstract}
Ultra-hot Jupiters (UHJs) are natural laboratories to study extreme physics in planetary atmospheres and their rich observational data sets are yet to be confronted with models with varying complexities at a population level. In this work, we update the general circulation model of Tan \& Komacek (2019) to include a non-grey radiative transfer scheme and apply it to simulate the realistic thermal structures, phase-dependent spectra, and wavelength-dependent phase curves of UHJs. We performed grids of models over a large range of equilibrium temperatures and  rotation periods for varying assumptions, showing that the fractional day-night brightness temperature differences remain almost constant or slightly increase with increasing equilibrium temperature from the visible to mid-infrared wavelengths. This differs from previous work primarily due to the increasing planetary rotation rate with increasing equilibrium temperature for fixed host star type. Radiative effects of varying atmospheric compositions become more significant in dayside brightness temperature in longer wavelengths.  Data-model comparisons of dayside brightness temperatures and phase curve amplitudes as a function of equilibrium temperature are in broad agreement.  Observations show a large scatter compared to models even with a range of different assumptions, indicating significantly varying intrinsic properties in the hot Jupiter population. Our cloud-free models generally struggle to match all observations for individual targets with a single set of parameter choices, indicating the need for extra processes for understanding the heat transport of UHJs. 
\end{abstract}

\begin{keywords}
hydrodynamics --- methods: numerical --- planets and satellites: atmospheres --- planets and satellites: gaseous planets 
\end{keywords}

\section{Introduction} \label{sec:intro}

Ultra-hot Jupiters are gas giant planets in extremely close-in orbits around their host stars, with  full-redistribution equilibrium temperatures in excess of 2200 K. The enormous stellar irradiation on their dayside places their atmospheres between the cooler hot Jupiters and those of late-type stars, making them ideal laboratories to understand the transition of stellar to planetary atmospheres and the atmospheric physics in extreme conditions. Owing to the ease of observability, blooming space-based \citep{kreidberg2018,Mansfield:2018aa,brightstar2023,mikal2022} and ground-based \citep{Ehrenreich:2020aa,Borsa:2021wr,2022A&A...668A.176P} observations over the past years have revealed the phase-dependent atmospheric conditions of these objects.

A notable difference between the atmospheres of hot and ultra-hot Jupiters is expected to be the thermal dissociation of molecules which could significantly affect the thermal structure and observations of the atmospheres \citep{kitzmann2018, lothringer2018, parmentier2018}. Key radiatively active molecules could be thermally dissociated and thus could impact the radiative balance and resulting emission spectra of these objects. Thermal dissociation of the major hydrogen molecules could significantly affect the dynamics of these atmospheres, determining the day-to-night heat transport efficiency and the detailed 3D circulation patterns \citep{bell2018,komacek2018rnaas,tan2019uhj,roth2021}. 

Comparisons between observations and general circulation models (GCMs) are vital to uncovering physical processes shaping the three-dimensional (3D) atmospheres of hot Jupiters. Two modeling strategies are commonly used in literature. The first approach is to model individual planets that have detailed observations and  models can be fine-tuned with different assumptions to  match many aspects of the observations including spectrum, the amplitudes of the phase curves, phase offsets, and the detailed phase-curve shapes. This approach has been conducted on many individual targets, revealing in-depth characterization of their individual atmospheres (e.g., \citealp{showman2009,zellem2014, stevenson2017spitzer, kreidberg2018, arcangeli2019,  steinrueck2019,lines2019,mansfield2020,carone2020,may2021,mikal2022,lee2022,beltz2022b,schneider2022exploring,zamyatina2023,lee2023mini}). 

Another approach, instead of focusing on explaining data for individual targets, aims to model  key trends as a function of system parameters and to identify the common physical mechanisms that possibly sculpt some key behaviors of the planetary population by systematic data-model comparisons over the parameter space. \cite{perna2012}, \cite{perezbecker2013}, and \cite{komacek2017} investigated the trend of day-night temperature variations under dynamical wave adjustment, radiative damping, and possible wind dissipation mechanisms, showing that the fractional day-night temperature variation is expected to increase with equilibrium temperature. \cite{tan2019uhj} modeled the ultra-hot Jupiter population using the updated GCM with the semi-grey radiative transfer and dynamical effects of hydrogen dissociation and recombination that are proposed to be important in ultra-hot Jupiters \citep{bell2018}, and showed that the fractional day-night temperature variation may no longer increase with equilibrium temperature in the ultra-hot regime. While these studies are illuminating in understanding the dynamical mechanisms, the simplified  radiative transfer calculations were not  sufficient to precisely model the atmospheric thermal structures and so insufficient to facilitate fair comparisons to phase-curve observations which are intrinsically wavelength dependent.  \cite{parmentier2016} and \cite{parmentier2021} carried out large grids of GCM simulations using the non-grey GCM originally developed in \cite{showman2009} and were able to pinpoint the roles of clouds and dynamics affecting the wavelength-dependent phase-curve observations of the canonical hot Jupiters population. This work is now being pursued by Roth et al. (submitted), which is looking at a grid of several thousand models. However, neither the dynamical core of \cite{parmentier2021} nor of Roth et al. (submitted) include the important effects of hydrogen dissociation and recombination and, therefore did not extend fully to the ultra-hot-Jupiter regime. In summary, there is now a gap in the GCM modeling of the ultra-hot Jupiter population that needs to be filled with more complete GCM simulations. 

In this work, using  GCMs with an updated non-grey radiative transfer scheme, we improve upon our previous work of \cite{tan2019uhj} and perform simulations over grids of varying equilibrium temperature to understand the trends of the day-night temperature variation of ultra-hot Jupiters. By comparing the modeling results to the rich observations of the ultra-hot Jupiter population, we wish to identify whether our model physics is adequate to explain the observed trends, or if not, what's likely missing in our models.  We use the SPARC/MITgcm, originally developed in \cite{showman2009}, and couple it with the hydrogen dissociation and recombination treatments in \cite{tan2019uhj} and \cite{komacek2022}. The realistic radiative heating and cooling rates in our updated models guarantee more precise modeling of the thermal structures, dynamics and chemistry, as well as more self-consistent comparisons to wavelength-dependent phase-curve observations.

Our work is also complementary to a variety of existing  GCM parameter surveys of hot Jupiter's atmospheres. Some of these parameter studies do not perform a systematic model-data comparison but  focus on the theoretical understanding of various physical and chemical processes over a large parameter space.  \cite{roman2021} included temperature-dependent clouds with  radiative feedback to assess the effects on phase curves as a function of equilibrium temperature. \cite{helling2022} computed cloud microphysics based on a large grid of GCM outputs, showing cloud  inhomogeneity over a wide range of planetary  parameters. \cite{baeyens2021} conducted a grid of GCM with a Newtonian cooling scheme and outputs are used to calculate chemical mixing, unveiling trends in the atmospheric chemistry of hot Jupiters. 

The outline of this work is as follows. Section \ref{sec:model} describes the setup of the GCM simulations, radiative transfer post-processing using \texttt{PICASO}, and system parameters. In Section \ref{sec:results}, we present results from the main suites of the model grids as a function of planetary equilibrium temperature assuming solar atmospheric metallicity, including the basic climatology, spectral, and phase-curve amplitude calculations. We summarize our key findings in Section \ref{sec:summary}. In the Appendix \ref{sec:metallicity}, motivated by the systematic mismatch of our results to the observed TESS  day-night temperature variations, we explore additional effects of varying atmospheric metallicity and carbon-to-oxygen ratio on the phase-curve amplitudes, using TOI 1431b as a template.

\section{Model Setup} \label{sec:model}

\subsection{The 3D general circulation model: \texttt{SPARC/MITgcm}}
The global atmospheric circulation and thermal structure of hot Jupiters were simulated using the SPARC/MITgcm \citep{showman2009}. The model solves the global primitive equations of dynamical meteorology on a cubed-sphere grid. This equation set is suitable to model the outermost observable layers of giant planets and brown dwarfs. Radiative heating and cooling rates in each column of the GCM are calculated using the non-grey radiative transfer model of \cite{marley1999} that solves the two-stream radiative transfer equations and employs the correlated-k method. The correlated-k opacity tables have been updated based on the opacities used in \cite{marley2021}. We assume thermochemical equilibrium in the atmosphere. In addition to molecular and atomic opacities in k coefficients listed in Table \ref{table:params}, the radiative transfer model also includes atomic line absorption from the neutral alkali metals, continuum opacity sources including pressure-induced opacities, bound-free and free-free opacity from ${\rm H}^-$ and ${\rm H_2}^+$ and free-free opacity from ${\rm H_2}^-$ and ${\rm He}^+$, as well as electron scattering and Rayleigh scattering from hydrogen and helium. See \cite{marley2021} for a detailed description. The k-coefficient tables used in the GCM  use 11 frequency bins as described in \cite{kataria2015}.

The hydrogen dissociation and recombination (${\rm H_2}$-${\rm H}$) is an important heating and cooling process in atmospheres of ultra-hot Jupiters \citep{bell2018,komacek2018rnaas}. We utilize  the same numerical schemes as idealized GCMs in \cite{tan2019uhj} that implement thermodynamically active tracers to track the evolution of atomic hydrogen mixing ratio. Our current implementation is the same as that in \cite{komacek2022} in which inert helium is included as the mass budget at Solar abundance. An alternative approach that does not need to invoke tracers is to cast the heating/cooling as a buffer effect which effectively increases the heat capacity at constant pressure when there is dissociation or recombination, as shown in a pseudo-2D framework of \cite{roth2021}. That approach assumes the instantaneous thermal equilibrium of hydrogen. In our tracer-based ${\rm H_2}$-${\rm H}$ scheme, we applied a ``soft" adjustment in which the conversion occurs over a finite timescale rather than instantaneously reaching local ${\rm H_2}$-${\rm H}$ equilibrium. This soft adjustment helps to maintain the numerical stability of our models by slightly reducing the extreme horizontal temperature gradients locally occurring in the model. When the ${\rm H_2}$-${\rm H}$ conversion timescale is very close to the dynamical time step (mimicking instantaneous adjustment), models become numerically unstable. 
The numerical challenge might stem from rapid dynamical instabilities associated with the large horizontal temperature and ${\rm H_2}$-H contrasts and they are difficult to be resolved within reasonable time steps.
After some tests, we choose a conservative ${\rm H_2}$-H relaxation timescale  of 40 s \footnote{ Realistic chemical calculations (\citealp{tsai2018toward}, Shang-Min Tsai, personal communication) suggest that, for example, at $10^{-3}$ bar, the expected chemical timescale of H$_2$ thermal dissociation is between $10^2$ to $10^3$ s at 2000 K and between 10 to $10^2$ s at 3000 K. Our choice of relaxation timescale = 40 s is reasonable at relatively low pressures. At $10^{-1}$ bar, the expected timescale is well below 1 s for temperatures higher than 1800 K. At relatively high pressures, the reaction should be considered instantaneous in GCM  step timescales. Dynamical timescales relevant in large scales can be estimated as $R/c$ where $R$ is the planetary radius and $c$ could be either horizontal wind speed or Kelvin wave phase speed (both are on the order of several ${\rm km~s^{-1}}$), and are typically on the order of $10^4$ to $10^5$ s. Our choice of chemical relaxation timescale = 40 s is much smaller than the dynamical timescales.  } (2-4 times the dynamical time step of the model) to guarantee stability and this is applied for all  models for consistency. Assuming a peak horizontal wind speed of $\sim10\;{\rm km\;s^{-1}}$, the minimum characteristic ``disequilibrium" length scale of hydrogen is  $10^5-10^6$~m which is much smaller than the planetary radius. Note that the atomic hydrogen mixing ratio in our tracer scheme which affects the dynamics and heating/cooling rates could be slightly different from those calculated by the full equilibrium chemistry calculation  used in the radiative transfer scheme, but this is a small trade-off in order to maintain numerical stability.

A Rayleigh drag with a coefficient of drag timescale of $10^5$~s at the bottom and linearly decreasing to zero with decreasing pressure until a pressure of 20 bars is applied in all models. This formulation is similar to that in \cite{liu2013} and mimics interactions with the interior. In practice, this ``basal'' drag scheme helps with model convergence and numerical stability. In addition, another drag with a globally uniform drag timescale $\tdrag$ is applied in a subset of the models to test the sensitivity of circulation and phase curves. In the main text, by "models without drag", we mean models without the uniform drag but still retaining the basal drag.

The temperature of the lowest model layer center (at a pressure of about 85 bars) in the GCM is relaxed towards a prescribed value of 3344 K over a timescale of $10^2$~s. This deep atmospheric profile is roughly consistent with those presented in \cite{Thorngren:2019aa} for inflated hot Jupiters. However, hot Jupiter inflation varies with planetary $\teq$ and the bottom temperature should in principle vary with $\teq$. Our prescribed value is somewhat conservative for ultra-hot Jupiters because we want to minimize the impact of the (uncertain) interior heat flux on the day-night temperature difference. This value is also held fixed in our grids to allow us to study the influence of only rotation period, host star type, and irradiation on the dynamics and heat transport. Our bottom boundary condition differs from previous setups of hot Jupiters (e.g., \citealp{showman2009}) in which a uniform upwelling heat flux was prescribed. Our bottom boundary condition mimics a rapid entropy mixing by vigorous interior convection. We argue that this is more appropriate here because convection is expected to homogenize the entropy (towards the uniform interior entropy) rather than heat flux, and the bottom of our domain is expected to lie below the radiative-convective boundary \citep{Thorngren:2019aa}. Using a similar hydrogen dissociation modeling framework as here, \cite{komacek2022the} recently showed that different bottom boundary treatments could result in differences in the isobaric temperature and circulation near the photosphere.

\subsection{Radiative transfer post-processing: \texttt{PICASO}}
Based on GCM outputs, we calculate the time-dependent planet spectra and observables using \texttt{PICASO} which is an open-source radiative transfer code for computing the reflected, thermal, and transmission spectrum of planets and brown dwarfs \citep{batalha2019,Mukherjee2023}.\footnote{Publicly available on Github \url{https://github.com/natashabatalha/picaso}} The methodology of \texttt{PICASO} to calculate the radiative transfer for thermal emission is the same as those in \cite{marley1999}. Emission, reflection spectra and phase curve calculations by \texttt{PICASO} based on 3D GCM outputs of hot Jupiters have been carried out in a few studies \citep{Adams2022,Robbins2022}.

\texttt{PICASO} has the same opacity source as the \texttt{SPARC/MITgcm} and this guarantees energetically consistent treatment between the GCM outputs and post-processing and can rigorously test effects of heat transport and ${\rm H_2}$-${\rm H}$ dissociation on the spectra and wavelength-dependent phase curves. In this work, we use the mode of correlated-k methods in \texttt{PICASO} with pre-calculated k-coefficients and equilibrium chemistry. These k tables have a much higher frequency resolution of  622 bins from about 0.26 to 267 $\mu$m than the version used in the GCM.
 
\begin{table}
\vspace{0.5cm}
\setlength{\tabcolsep}{0pt}
\footnotesize
\begin{center}
\begin{tabular}{ l l }
\hline
{\bf Parameter} & \quad {\bf Value} \\
\hline
Stellar effective temperature &\quad [5500, 6000, 6500] K \\
Stellar mass &\quad [0.94, 1.08, 1.34] $M_{\rm sun}$ \\
Stellar radius &\quad [0.91, 1.10, 1.50] $R_{\rm sun}$ \\
Planet radius  &\quad  $10^8 {\rm m}$ \\
Planet gravity  &\quad 10 m~s$^{-2}$\\
Planet equilibrium temperature & \quad [1800, 2000, 2200, 2400, 2600] K \\
Planet rotation period  &\quad  [1.05, 0.77, 0.58, 0.44, 0.35] days \\
&\quad [1.69, 1.23, 0.93, 0.71, 0.56] days \\
&\quad [3.07, 2.24, 1.68, 1.30, 1.06] days \\
Temperature at 85 bars &\quad 3344  K \\
H tracer relaxation timescale &\quad 40 s \\ 
Drag timescale ($\tau_\mathrm{drag}$) &\quad $\infty$, $10^4$ s\\
Metallicity and C/O ratio &\quad $1\times$ solar\\
Species in k-coefficient tables & \quad ${\rm H_2O}$, ${\rm CO_2}$,${\rm CO}$, ${\rm CH_4}$, ${\rm NH_3}$, ${\rm OCS}$,  \\
& \quad ${\rm N_2}$, ${\rm HCN}$, ${\rm C_2H_2}$, ${\rm C_2H_6}$, ${\rm H_2S}$,   \\
& \quad   ${\rm C_2H_4}$,  ${\rm FeH}$, ${\rm CrH}$, ${\rm LiCl}$,  ${\rm SiO}$, ${\rm MgH}$, \\
& \quad  [ ${\rm TiO}$, ${\rm VO}$, ${\rm Fe}$, ${\rm H_2}$, ${\rm H_3^+}$,  ${\rm LiH}$, ${\rm LiF}$ ]\\
Horizontal resolution &\quad C32, C48 \\
Vertical resolution &\quad 53 layers\\
Lower boundary &\quad 100 bars \\
Upper boundary &\quad $5\times10^{-6}$ bar \\
Shapiro filter order &\quad  4 \\
Shapiro filter timescale &\quad 50 s \\ 
Dynamical time step &\quad 10 - 20 s\\
Radiative time step &\quad 50 s \\
\hline
\end{tabular}%
\end{center}
\caption{Stellar, planetary, and numerical parameters assumed for the suite of GCM simulations presented Section \ref{sec:results}. The last bracketed line of species in k-coefficient tables indicates parameter sensitivity tests over a subset of models in which these species are not included in the radiative transfer. }
\label{table:params}
\end{table}

\subsection{Parameters of our grid simulations}
Following up our previous approach using idealized GCMs \citep{tan2019uhj} and many other GCM studies, we sweep a wide parameter space in order to understand the trend in the ultra-hot Jupiter population.

We specify a stellar effective temperature, mass, and radius, and then a set of orbital periods can be determined for a set of zero-albedo planetary equilibrium temperatures. We assume three stellar effective temperatures $T_{\rm star}$ of 5500 K, 6000 K, and 6500 K which should be representative of most stellar hosts of hot Jupiters. Using  stellar mass-radius-temperature relations found by \cite{eker2018} based on solar neighbourhoods, we set the stellar masses $M_{\rm star}$ as 0.94$M_{\rm sun}$, 1.08$M_{\rm sun}$ and 1.34$M_{\rm sun}$, and the stellar radius $R_{\rm star}$ as 0.91$R_{\rm sun}$, 1.1$R_{\rm sun}$ and 1.5$R_{\rm sun}$, respectively. With a given planetary equilibrium temperature $T_{\rm eq}$ assuming zero albedos, the orbital period  (therefore the rotation period $P_{\rm rot}$ assuming tidal locking) is given by
\begin{equation}
    P_{\rm rot}=\frac{\pi}{\sqrt{2}}\left(\frac{T_{\rm star}}{T_{\rm eq}}\right)^3\sqrt{\frac{R^3_{\rm star}}{GM_{\rm star}}},
\end{equation}
where $G$ is the gravitational constant. Within the planetary $\teq$ that we consider in this work, tidal synchronization is a reasonable assumption \citep{guillot1996}.  As will be discussed below, the change of rotation rate of tidally locked planets corresponding to different stellar properties and $T_{\rm eq}$ is the dominant effect on the circulation and day-to-night heat transport, rather than the nature of the incident stellar spectra. 

We model atmospheres with equilibrium temperatures from 1800 to 2600 K, encompassing the majority of the ultra-hot Jupiter population.  The sets of rotation period for planetary equilibrium temperature of [1800, 2000, 2200, 2400, 2600] K for three stellar temperatures are [1.05, 0.77, 0.58, 0.44, 0.35] days, [1.69, 1.23, 0.93, 0.71, 0.56] days and [3.07, 2.24, 1.68, 1.3, 1.06] days, respectively with increasing stellar effective temperature. The horizontal resolution of most simulations is C32, equivalent to about $2.81^{\circ}\times 2.81^{\circ}$ in longitude and latitude; some rapidly rotating, drag-free cases of $\teq=2200, 2400$ and 2600 K with $T_{\rm star}=5500$, and $\teq=2600$ K with $T_{\rm star}=6000$, use a C48 resolution (about $1.88^{\circ}\times 1.88^{\circ}$ in longitude and latitude) to better resolve smaller horizontal structures.

In the main set of simulations of Section \ref{sec:results}, we include key optical absorbing agents including TiO, VO and Fe, which have been known to drive a strong thermal inversion on the dayside atmosphere and can be detected on ultra-hot Jupiters with high resolution spectroscopy \citep{Hoeijmakers:2018aa,Hoeijmakers:2020aa,Ehrenreich:2020aa,Kesseli:2021uk,2023arXiv230608739P}. 
Models without drag are integrated for 1350 to 2250 days, and models with $\tdrag=10^4$~s are integrated for 625 to 1000 days, depending on the  timescale for each case to reach an energy equilibrium. Outputs of the last 200 days time-averaged for the results shown in Section \ref{sec:results}. Key parameters of our models are summarized in Table \ref{table:params}.


\begin{figure*}      
\centering
\includegraphics[width=2\columnwidth]{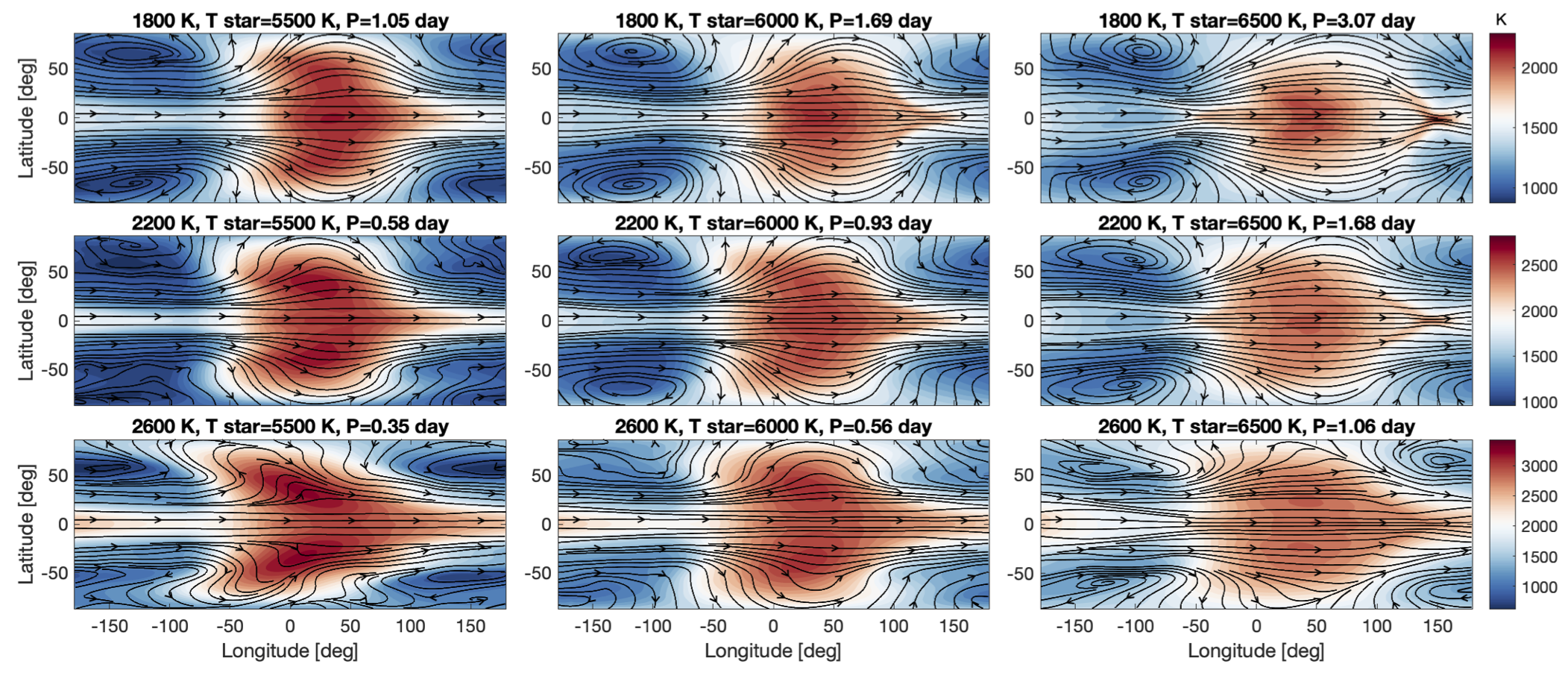}
\caption{Temperature maps at 55 mbar for drag-free models with varying planetary equilibrium temperature from 1800 K to 2600 K (from top to bottom) and  varying  tidally synchronized rotation periods implied by stellar effective temperatures of 5500 K, 6000 K and 6500 K (from left to right). Horizontal winds are represented by streamlines and arrows which are normalized separately for each panel. The rotation period of each model is labeled above each panel. These results were obtained by time-averaging over the last 200-day drag-free model outputs. Temperature color maps of the same planetary equilibrium temperature (e.g., in each row) share the same color bar on the right.
}
\label{fig.temperaturemap24}
\end{figure*} 

\begin{figure*}      
\centering
\includegraphics[width=2\columnwidth]{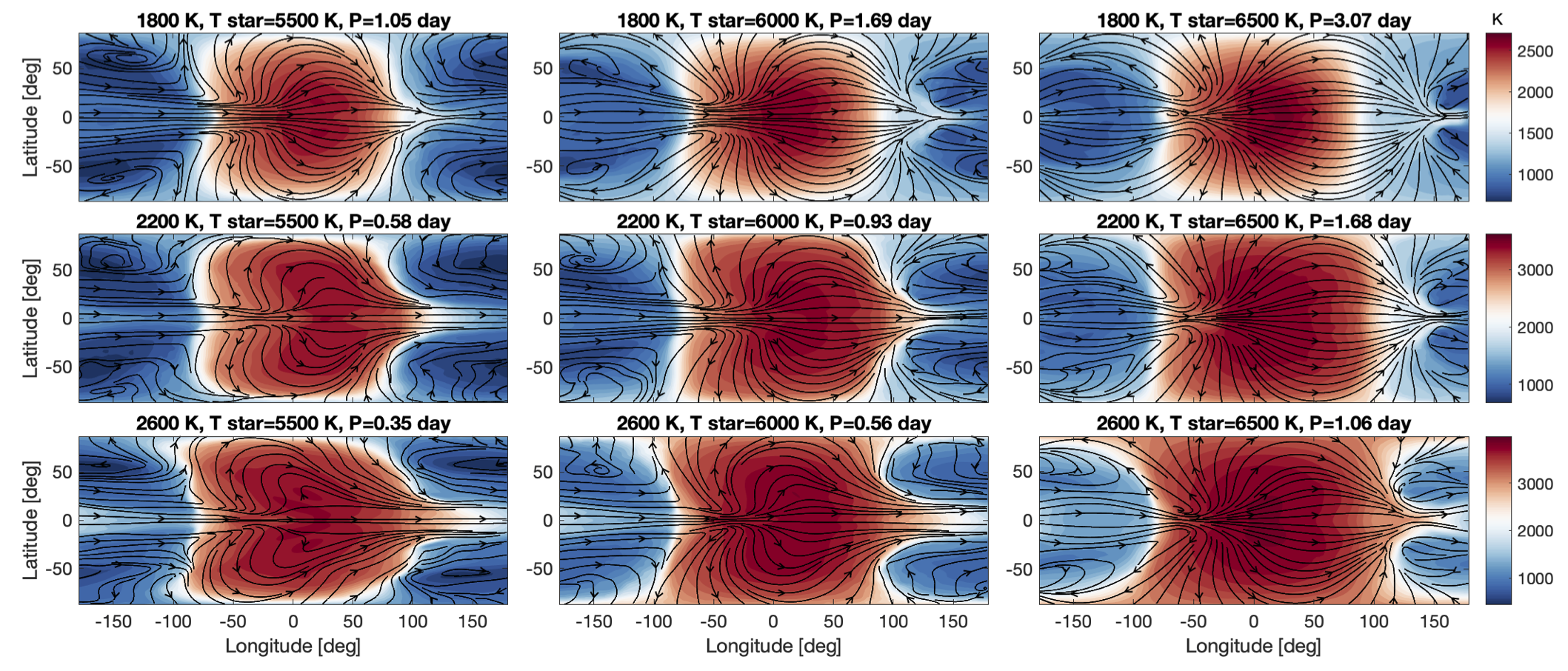}
\caption{Same as Figure \ref{fig.temperaturemap24} but showing temperature maps at 0.6 mbar.
}
\label{fig.temperaturemap40}
\end{figure*}

\begin{figure*}      
\centering
\includegraphics[width=2\columnwidth]{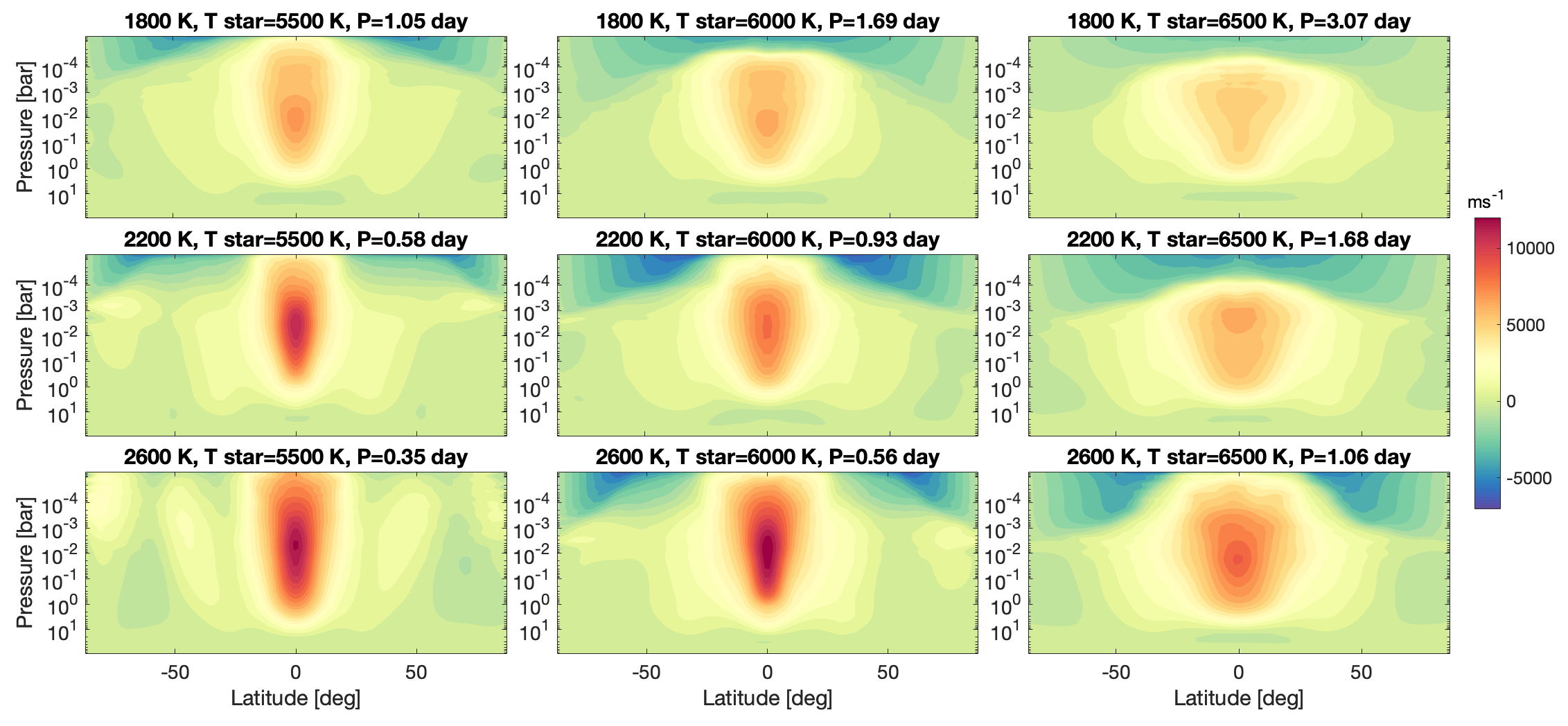}
\caption{Zonal-mean zonal winds for drag-free models with varying planetary equilibrium temperature from 1800 K to 2600 K (from top to bottom) and varying stellar effective temperature of 5500 K, 6000 K, and 6500 K (from left to right). The rotation period of each model is labeled above each panel. These results were obtained by time-averaging over the last 200-day drag-free model outputs.
}
\label{fig.jet}
\end{figure*} 

\section{Results of the main suits of models} \label{sec:results}
\subsection{Basic circulation patterns and spectra}
\label{sec:basic}
This subsection describes the temperature maps, zonal jets, dayside and nightside temperature-pressure (TP) structures, effect of ${\rm H_2}$-${\rm H}$ dissociation on heat transport, and the dayside and nightside thermal spectra resulting from the circulation. These results are based on the main suites of simulations assuming 1x solar compositions described in Table \ref{table:params}.

\subsubsection{Temperature fields and jets}
Figure \ref{fig.temperaturemap24} shows the horizontal temperature maps at 55 mbar which is near the thermal photosphere level for a subset of the models without drag. Equilibrium temperature increases from 1800 K to 2600 K from the top to the bottom, and stellar temperature increases from 5500 K to 6500 K from left to right. Streamlines with arrows represent horizontal velocity fields. All models show qualitatively consistent features,  including a prominent equatorial eastward jet and stationary global-scale waves which are shifted eastward relatively to the sub-stellar longitude of $0^{\circ}$ and extend from the equator to the poles. These are the canonical circulation patterns that emerge from the strongly irradiated, moderately rotating, tidally locked hot Jupiter dynamical regimes (e.g., \citealp{showman2002,showman2009,heng2011, rauscher2012, mayne2014,cho2015,mendonca2016,Hammond:2018aa,tan2019uhj,carone2020,lee2021,hammond2021}, see also reviews by \citealp{heng2015}, \citealp{zhang2020}, and  \citealp{showman2020}). At a given stellar temperature, the extent to which the global thermal pattern is shifted eastward of the sub-stellar longitude lessens with increasing equilibrium temperatures. This is due to both the increasing radiative heating efficiency and the stronger rotation, both of which act against horizontal heat transport \citep{showman2013,tan2020wdbd}. This trend is the same for all three stellar temperatures.

At a lower pressure of 0.6 mbar as shown in Figure \ref{fig.temperaturemap40}, the dayside  atmospheres are hotter due to the strong thermal inversion caused by optical absorbers, and the atomic hydrogen mixing ratio is at higher values due to thermal dissociation. This larger sub-stellar-to-limb atomic hydrogen contrast results in a higher horizontal heat transport efficiency within the dayside from the heating due to hydrogen recombination on the nightside and limb and cooling due to dissociation near the substellar point \citep{tan2019uhj}. Compared to those at 55 mbar, the hot area is more significantly filled in the dayside. Furthermore, for the hotter cases, the hot areas extend substantially to the nightside through the circulation mostly at high latitudes. The extent to which the  pattern is shifted eastward of the sub-stellar longitude is weaker than those shown at 55 mbar, similar to that shown for canonical hot Jupiters \citep{showman2013b,lee2021}.  Rotation is vital to determine where the heat is preferentially transported. For example, on the left column where the rotation periods are shorter for a given equilibrium temperature, flows at high latitudes are strongly confined by the Coriolis force and form Rossby  gyres, whereas on the right column where rotation periods are longer, flows are less confined by the Coriolis force, with a weaker Rossby gyre and greater dayside to nightside (divergent) flow \citep{hammond2021} which efficiently warms the high latitude regions on the nightside. 

 All models show a dominant equatorial superrotating jet with speeds up to about 10 ${\rm km~s^{-1}}$ and most of the jets extend from nearly the model top boundary down to almost 10 bars where the basal drag starts to dissipate the horizontal winds. Figure \ref{fig.jet} shows the time-averaged zonal-mean zonal winds for the same set of models shown in Figure \ref{fig.temperaturemap24}. In models with a fixed stellar temperature, the jet speed increases with increasing planetary equilibrium temperature; in models with a specific planetary equilibrium temperature, the jet speed decreases with increasing stellar temperature (and hence increasing rotation period). These trends make sense intuitively. The first trend emerges likely because the eddy velocities are expected to increase with increasing equilibrium temperature in the absence of strong dissipation \citep{perezbecker2013,komacek2016,koll2018}; the superrotating jet is driven by horizontal convergence of the eddies \citep{showman2011}, therefore it is suspected that the jet speed is also positively correlated with the eddy magnitude which has been shown by numerical simulations and scaling analysis \citep{zhang2017, hammond2020}. On the second trend, given the same equilibrium temperature (hence the same expected eddy magnitude), the magnitude of eddy convergence is expected to increase when the eddy meridional lengthscale is smaller which is expected for faster rotators \citep{showman2011,pierrehumbert2019}. 

The meridional width of the jets was expected to decrease with decreasing rotation period due to the shrinking meridional extent of the eddies that is proportional to the equatorial deformation radius $L_d = \sqrt{\frac{c}{\beta}}$, where $c$ is the gravity wave phase speed, $\beta=df/dy$ at the equator, $f=4\pi \sin \phi / P$ is the Coriolis parameter, $P$ is the rotation period and $\phi$ is latitude.\footnote{The meridional extent of the eastward acceleration on the zonal flow by the stationary Matsuno-Gill waves could be modified by the presence of a background equatorial jet, see Figure 7 of \cite{Hammond:2018aa}. Therefore, one might wonder what wave speed is relevant -- the one Doppler shifted by a background jet or in a rest background (i.e., with no Doppler shift). Regardless of which wave speed should be used, numerical results of the jet widths in the moderate-to-fast rotating regime from \cite{tan2020wdbd} are well matched to the scaling of  rotation period.} Therefore, we would expect that the jet width is proportional to $\sqrt{P}$. In GCM simulations for both the non-synchronously rotating hot and warm Jupiters \citep{showman2015} and synchronously rotating close-in giant planets \citep{tan2020wdbd}, the decreasing meridional width of the jets with decreasing rotation periods are both investigated. In the rapid-to-moderate rotating regime, the meridional jet width showed a clear linear scaling to the equatorial deformation radius which is proportional to $\sqrt{P}$  when everything else is held fixed  \citep{tan2020wdbd}. Indeed, at the same $\teq$ but varying stellar temperatures and rotation periods (horizontal comparisons in Figure \ref{fig.jet}), we see this qualitative trend of increasing meridional jet with increasing rotation period. 

However, this is not the case in the vertical comparisons of Figure \ref{fig.jet} in which the jet meridional widths change little with varying rotation periods under a constant $T_{\rm star}$.    The vertical structure of the eddies might be somewhat sensitive to the heating and cooling rates which could affect their interactions with the mean flow \citep{tsai2014}.  A key varying factor in the vertical comparisons of Figure \ref{fig.jet} may be the change of eddy structures along with the change of effective temperature, and the wave speed $c$ associated with the eddies is no longer invariant.  Holding the opacity structure and rotation rate the same, \cite{tan2019uhj} showed that the eastward equatorial jets near the photosphere become weaker and slightly narrower as $\teq$ increases, and in addition, the upper layers transition to westward zonal-mean equatorial flow; however, we don't see this behavior in our new sets of models with more realistic heating/cooling rates and varying rotation. The sensitivity of the circulation and jet to heating and cooling rates have been shown by \cite{lee2021} and \cite{steinrueck2023}. In addition to the radiative heating and cooling rates, heating rates caused by ${\rm H_2}$-${\rm H}$ also change with equilibrium temperature, further complicating the eddy structure. Overall, the jet structure (determined by the wave-mean-flow interaction mechanism) requires a more thorough understanding when more realistic factors are considered.


\begin{figure*}      
\centering
\includegraphics[width=2\columnwidth]{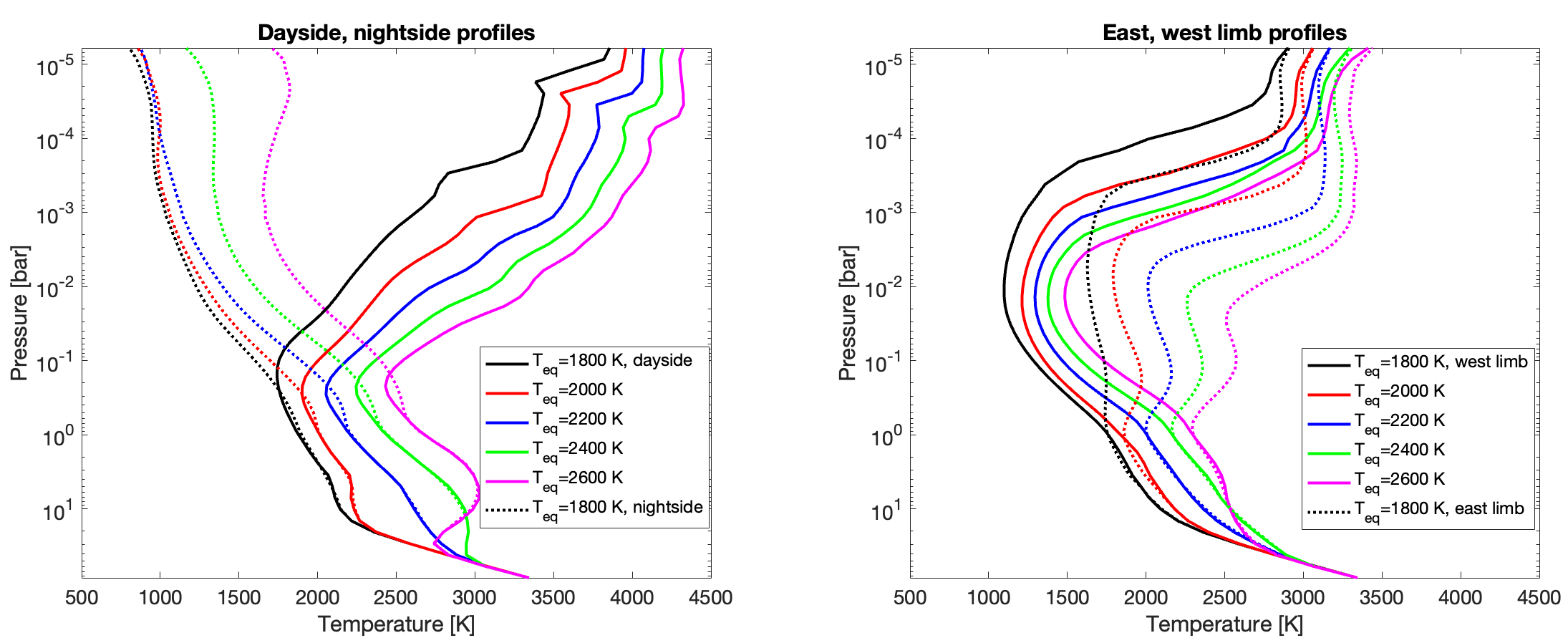}
\caption{Left panel: temperature-pressure profiles from the sub-stellar region (solid lines) and  the anti-sub-stellar region (dotted lines) from models with $\teq$ from 1800 K to 2600 K and without drag. Different  colors represent different planetary equilibrium temperatures.  Profiles near the sub-stellar area were averaged within $\pm20^{\circ}$ in longitude and latitude (where $0^{\circ}$ is the sub-stellar point); profiles near the anti-sub-stellar point were averaged over longitudes >$160^{\circ}$ and <$-160^{\circ}$ and latitudes within $\pm20^{\circ}$. Right panel: profiles averaged over the west limbs (solid lines) and east limbs (dashed lines).  Profiles of the west limbs were averaged over longitudes >$-100^{\circ}$ and <$-80^{\circ}$ and latitudes within $\pm75^{\circ}$, and profiles of the east limbs were averaged over longitudes >$80^{\circ}$ and <$100^{\circ}$ and latitudes within $\pm75^{\circ}$. The ``opening angles" in the transit observations should depend on $\teq$ \citep{caldas2019,wardenier2022all}, and here we fix the sampling longitudinal regions of the limb profiles merely for a demonstration of the heat transport effects. These results were obtained by time-averaging over the last 200-day simulations. The stellar effective temperature of these models is 6500 K.}
\label{fig.tps}
\end{figure*}



\begin{figure*}      
\centering
\includegraphics[width=2\columnwidth]{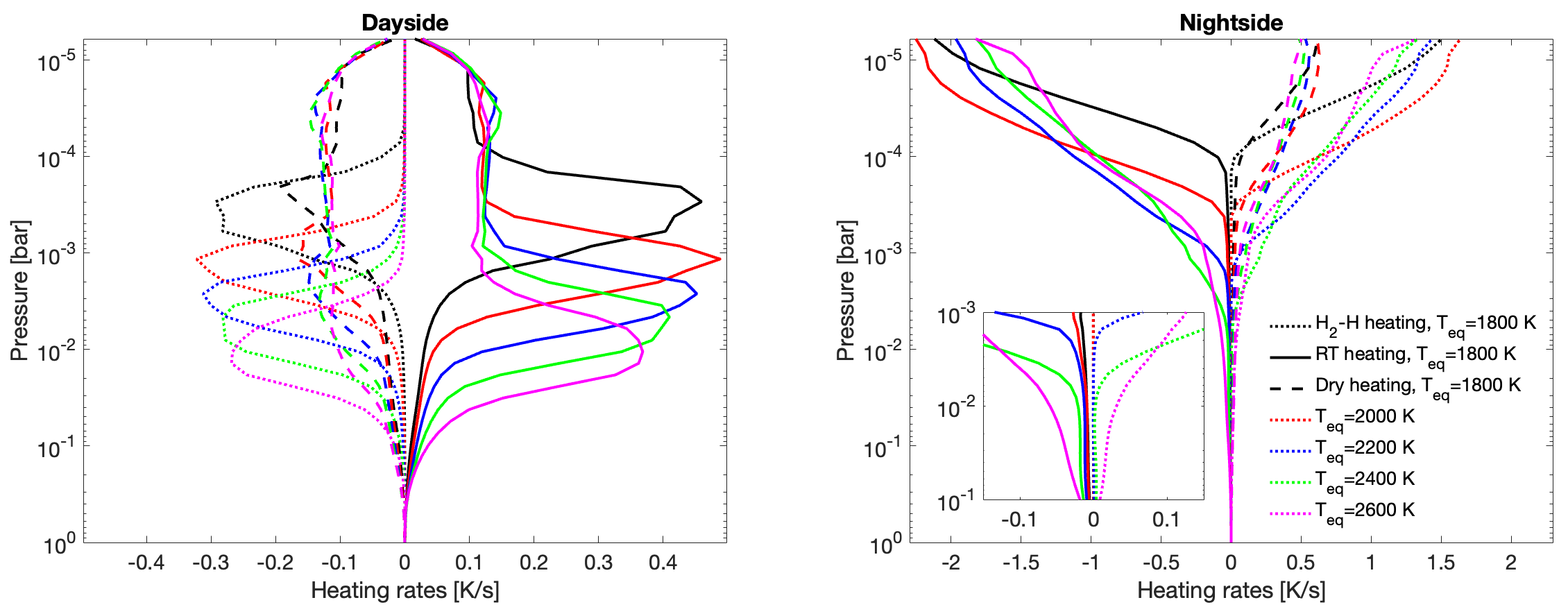}
\caption{Radiative heating and cooling profiles (solid lines) and ${\rm H_2}$-${\rm H}$ heating and cooling profiles (dotted lines) as a function of pressure for the set of models shown in Figure \ref{fig.tps}. Dashed lines are the ${\rm H_2}$-${\rm H}$ heating minus the radiative heating rates, and correspond mostly to the dynamical heating rates by the dry components.  The left panel contains the dayside profiles averaged over a domain that is $[\pm 60^{\circ}, ~\pm 60^{\circ}]$ of longitude and latitude around the sub-stellar point, and those in the right panel are similarly averaged over a domain around the anti-sub-stellar point.}
\label{fig.heatingrate}
\end{figure*}

\begin{figure*}      
\centering
\includegraphics[width=1.8\columnwidth]{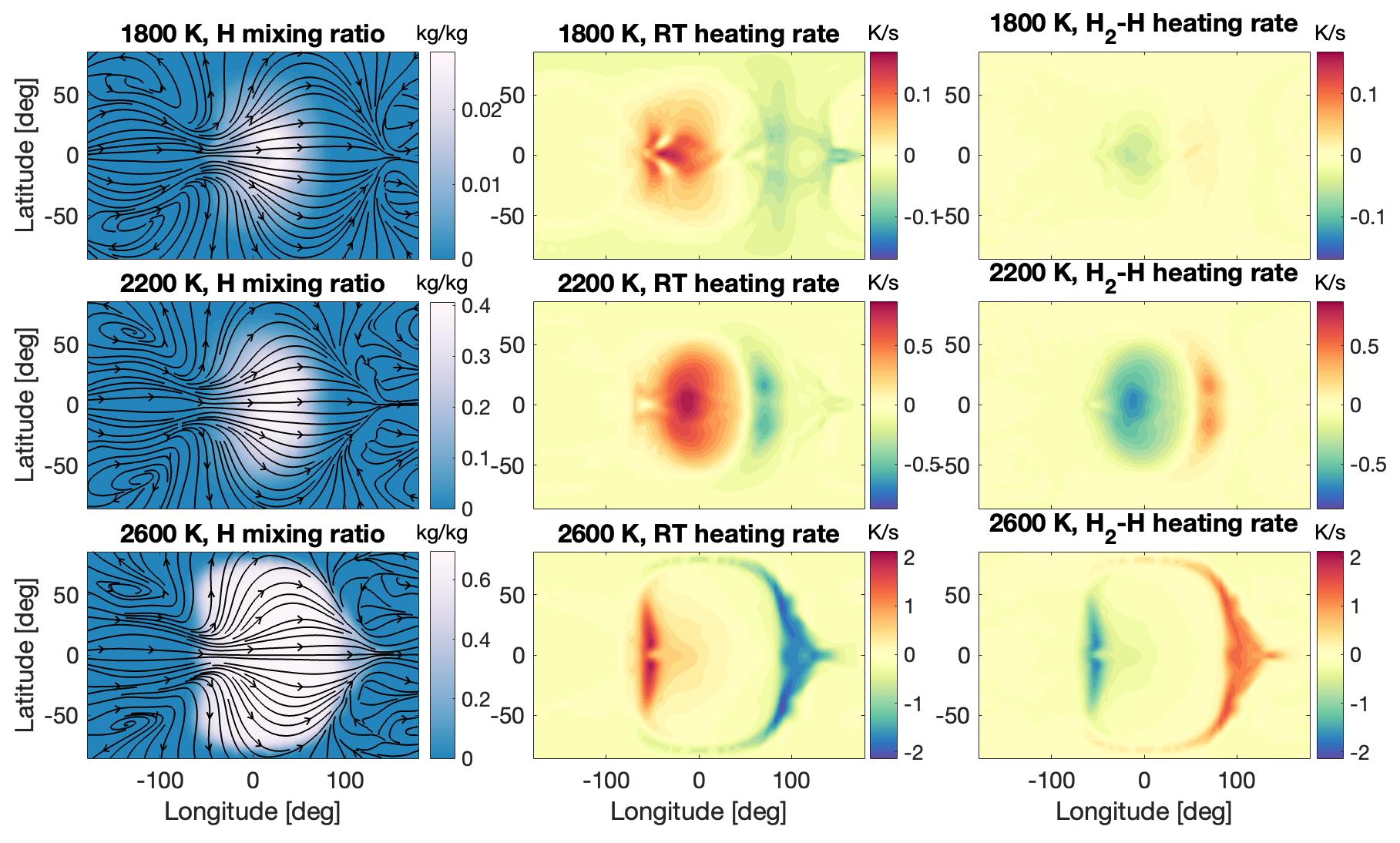}
\caption{ Isobaric maps at about 6 mbar for atomic hydrogen mass mixing ratio (left), radiative heating and cooling rates (middle) and ${\rm H_2}$-${\rm H}$ heating and cooling rates (right) as a function of longitude and latitudes. Streamlines with arrows on the left represent isobaric winds. Results are from models with $\teq=1800$, 2200 and 2600 K from top to bottom. The host star has an effective temperature of 6500 K. }
\label{fig.heatingratemap}
\end{figure*}

 \subsubsection{Temperature and heating rate profiles}
The left panel of Figure \ref{fig.tps} shows the TP profiles averaged over a small area around the sub-stellar point (solid lines) and around the anti-sub-stellar point (dotted lines) for a set of simulations with $\teq =1800$ to 2600 K, a stellar temperature of 6500 K, and without drag. The dayside thermal inversion above about 0.1 bar, i.e., temperature increases with increased altitude, is primarily driven by absorption of TiO and VO first at relatively high pressure and then by absorption of Fe at low pressures \citep{lothringer2018, parmentier2018,Hoeijmakers:2018aa}. Radiative cooling is efficient on the nightside to cool the atmosphere and the TP profiles show decreasing temperatures with decreased altitude on the nightside photospheres.  The overall nightside temperature variations among  $\teq=1800$ to 2200 K are small, and this has been discussed in \cite{parmentier2021} regarding radiative-dynamical timescale arguments. The nightside temperatures with $\teq>2200$ K show stronger $\teq$ dependence, presumably related to the ${\rm H_2}$-${\rm H}$ heating transport.

Dynamical heat transport induces strong thermal inversion and east-west asymmetry near the terminators. The temperature profiles averaged in the east and west limbs are shown in the right panel of Figure \ref{fig.tps}. Similar thermal inversion in the upper atmosphere near the terminators has been seen in other GCMs of ultra-hot Jupiters \citep{lee2022}, and the effect here is perhaps even stronger because of the additional ${\rm H_2}$-${\rm H}$ heat transport.  More interestingly, in the drag-free cases, the east-limb profiles are systematically hotter than the west-limb profiles, at some pressure levels showing east-west-limb temperature differences of more than 1000 K. The overall east-west-limb temperature differences decrease with decreasing $\teq$. In terms of pressure dependency, the east-west-limb differences also decrease with decreasing pressures in the upper layers; this happens at about $3.5\times 10^{-3}$ bar at $\teq=2600$ K and is at lower pressures for cooler cases (about $2\times 10^{-4}$ bar at $\teq=1800$ K). The pressure-dependent east-west-limb differences are partly related to the relatively stronger day-to-night circulation at low pressures shown in Figures \ref{fig.temperaturemap24} and \ref{fig.temperaturemap40}.  The east-west limb differences should be readily detectable in the transit lightcurve asymmetry by JWST  as has been demonstrated by recent work for WASP-39b (JWST Transiting Exoplanet Early Release Science Team, in prep.) as well as high-resolution transmission spectroscopy (e.g., \citealp{savel2023}).

With a correlated-k radiative transfer scheme which calculates realistic radiative heating and cooling rates and  TP structure, we can investigate pressure-dependent ${\rm H_2}$-${\rm H}$ heat transport, whereas we were not confident to do so in previous work with an idealized, semi-grey scheme \citep{tan2019uhj}. The efficiency of hydrogen heat transport relative to the nominal heat transport by the ``dry" atmosphere can be shown by the ${\rm H_2}$-${\rm H}$ heating rate profiles and  compare to the radiative heating profiles. Figure \ref{fig.heatingrate} shows the  radiative heating profiles (solid lines) and ${\rm H_2}$-${\rm H}$ heating profiles (dotted lines) for the set of models shown in Figure \ref{fig.tps}. The left panel contains the dayside profiles averaged over a domain that is $[\pm 60^{\circ}, ~\pm 60^{\circ}]$ of longitude and latitude around the sub-stellar point, and the right panel is similarly averaged over a domain around the anti-sub-stellar point. 

On the dayside, ${\rm H_2}$-${\rm H}$ cooling profiles are  small at both high and low pressures and show monotonic peaks at altitudes of about 10 mbar to 0.4 mbar depending on the model $\teq$. The small ${\rm H_2}$-${\rm H}$ cooling rates  at high pressures are because of the globally small atomic hydrogen fraction there. The small ${\rm H_2}$-${\rm H}$ rates at very low pressure are  because of the dissociation saturation, i.e., hydrogen is almost fully dissociated over the whole dayside region where the heating profiles are sampled, and  there are little regions with  large gradients of the atomic hydrogen radio either in the horizontal or vertical directions. The ${\rm H_2}$-${\rm H}$ heat transport  depends on the gradient of the atomic hydrogen mixing ratio rather than just the ratio itself, and so the ${\rm H_2}$-${\rm H}$ heat transport is halted at low pressures. Only in the intermedium pressure regions which are also  where the thermal photospheres roughly are, the transitions from molecular to atomic hydrogen lead to  significant ${\rm H_2}$-${\rm H}$ cooling that almost balances the radiative heating. The radiative heating profiles on the dayside show corresponding shapes to the hydrogen dissociation cooling profiles, but at low pressure, they are still significantly nonzero and are balanced by the cooling induced by the ``dry" atmospheric heat transport. The ``dry" heat transport cooling rate profiles are represented as the dashed lines in Figure \ref{fig.heatingrate}. Outside of regions where the radiative heating rates peak and are balanced largely by ${\rm H_2}$-${\rm H}$ cooling, the dry heating rates are the dominant balance to radiative heating. 

On the much cooler nightside, the atomic hydrogen mixing ratio  is generally smaller than on the dayside. For the hottest models,  the ${\rm H_2}$-${\rm H}$ heating rates near the thermal photosphere are comparable to that on the dayside. All models show strong  ${\rm H_2}$-${\rm H}$ heating at very low pressures above the thermal photosphere (the right panel of Figure \ref{fig.heatingrate}) but their effects should be invisible in the (low-resolution) emission spectrum. Likewise,  the radiative cooling profiles have similar shapes to those of  ${\rm H_2}$-${\rm H}$ heating but with slightly larger amplitudes.

 Figure \ref{fig.heatingratemap} displays isobaric maps at about 6 mbar for atomic hydrogen mixing ratio, radiative heating rates and H$_2$-H heating rates of models with $\teq=1800$, 2200 and 2600 K, for a closer look at the H mixing ratio and heating rate structures.  The shapes of the radiative heating rates, either as shown in the mean vertical profiles in Figure \ref{fig.heatingrate} or in horizontal maps, display a good correlation to the ${\rm H_2}$-${\rm H}$ heating rates but the magnitude of the latter is always smaller. This is expected when the hydrogen is near equilibrium. Neglecting time dependency and frictional dissipation of heat, the thermodynamic equation can be written as (Equation [17] in \citealp{tan2019uhj})
\begin{equation}
    (\mathbf{v}\cdot \nabla_p+\omega \frac{\partial}{\partial p})(\bar{c}_pT+\mathcal{L}_h q) - \frac{\omega}{\bar{\rho}} = g\frac{\partial F}{\partial p},
    \label{eq.thermo}
\end{equation}
 where $\mathbf{v}$ is the isobaric wind vector, $p$ is pressure, $T$ is temperature, $q$ is mass mixing ratio of atomic hydrogen H relative to total mass,  $\omega$ is the vertical velocity at pressure coordinate, $\nabla_p$ is the horizontal gradient at pressure coordinate, $\bar{c}_p$ is the mean specific heat at constant pressure, $\mathcal{L}_h$ is the latent heat associated with hydrogen dissociation, $\bar{\rho}$ is the mean gas density, and $F$ is the radiative energy flux.  
If horizontal  transport is the dominant process, the above equation is simplified as
\begin{equation}
    \mathbf{v}\cdot (\bar{c}_p+\mathcal{L}_h \frac{\partial q}{\partial T}) \nabla_p T \sim g\frac{\partial F}{\partial p}.
    \label{eq.thermo2}
\end{equation}
Heat transport is done by the ``dry" and ``latent heat" components which are represented by the $\bar{c}_p$ and $\mathcal{L}_h \frac{\partial q}{\partial T}$ (equivalently, the latter can be seen as a buffer in the heat capacity, \citealp{roth2021}). In nearly chemical equilibrium, i.e., $\frac{\partial q}{\partial T}\approx \frac{\partial q_{\rm eq}}{\partial T}$,  $\mathcal{L}_h \frac{\partial q_{\rm eq}}{\partial T}$ at the isobaric surface is a function of temperature alone and is usually a factor of a few larger than  $\bar{c}_p$ at relatively high temperature  before saturation. Radiative heating is mostly balanced by the latent heating in these regions, therefore showing good spatial correlations between radiative and latent heating.

\begin{figure*}      
\centering
\includegraphics[width=1.9\columnwidth]{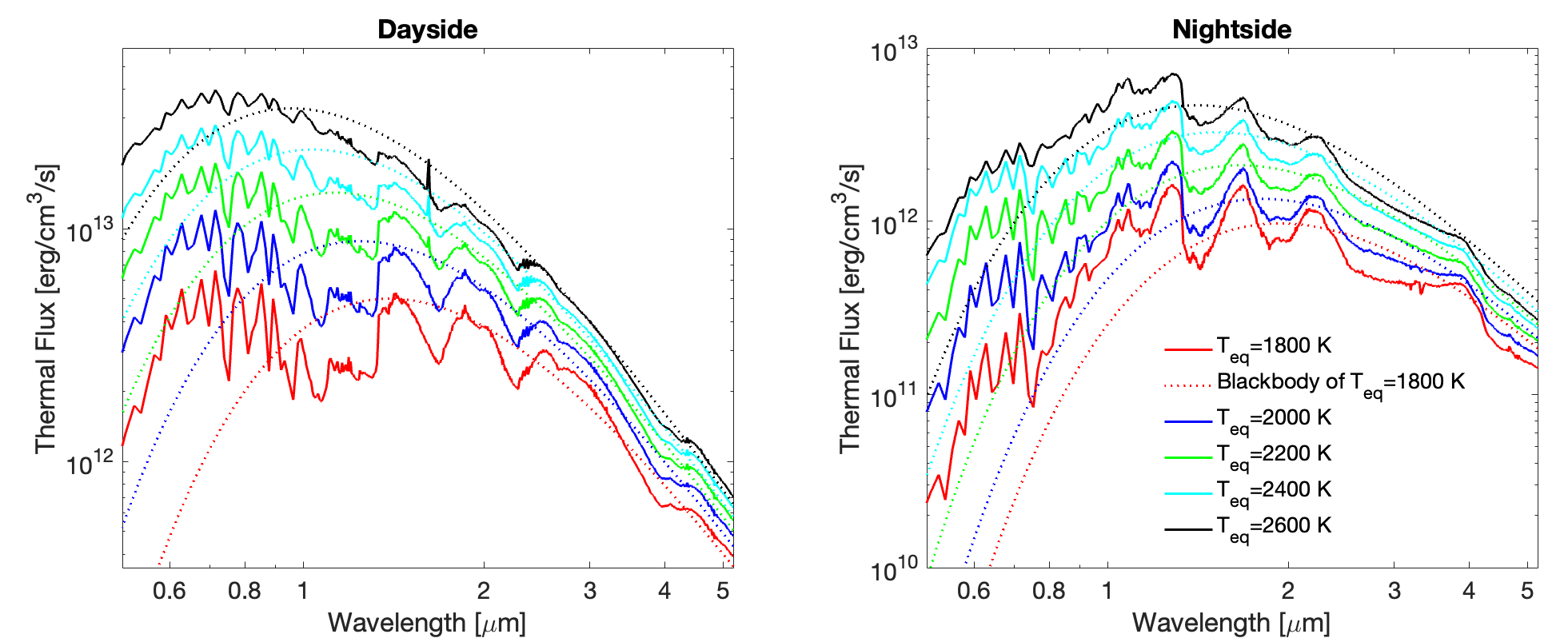}
\caption{Thermal emission spectra (solid lines) of models with different equilibrium temperatures and a stellar  temperature of 6500 K on the dayside (left) and on the nightside (right). The models shown here are the same as those shown in Figure \ref{fig.tps}. The dotted lines are Blackbody spectra with total energy corresponding to those of the model output spectra integrated over the full wavelength range; they aid visualization of the spectral distributions.
}
\label{fig.spectra-dragfree}
\end{figure*}

\subsubsection{Emission spectra}

GCM outputs of the last 200 days were time-averaged and fed into the radiative transfer code \texttt{PICASO} to post-process thermal emission spectra as a function of phase as the planet rotates. Instantaneous time variability during the orbital motion is therefore not considered in our post-processing. 

Figure \ref{fig.spectra-dragfree} shows the dayside (left) and nightside (right) thermal emission spectra for the same suite of models with a stellar temperature of 6500 K shown in Figure \ref{fig.tps}. Spectra on the dayside all show emission features that are associated with the strong thermal inversions on the dayside, especially at wavelength regions sculpted by the TiO, H$_2$O, and CO molecules. The strength of the features progressively decreases  with increased temperature of the atmosphere, primarily due to the thermal dissociation of the key molecules at high altitudes and therefore the decreasing molecular scale heights probed by the emission (e.g., \citealp{parmentier2018}). Interestingly, the sharp and narrow emission features at about 1.6 $\mu$m are caused by the CO molecules, but this feature is only prominent at high temperatures. This is because CO molecules are more resistant to higher temperatures compared with  H$_2$O, and its' spectral features at 1.6 $\mu$m appears more robustly when H$_2$O becomes significantly dissociated at high temperatures ($\gtrapprox 3000$ K). On the nightside, the near-IR and IR spectral features change to absorption because of the decreasing temperature with increasing altitude (see Figure \ref{fig.tps}). The feature strength also slightly decreases with increased temperature but is not as obvious as that on the dayside. The nightside spectra exhibit significant excesses of fluxes in the visible wavelength compared to the Blackbody spectra, and those are emitted mostly from the terminator regions when the planetary nightsides face toward the observer according to the spatially resolved flux map (not shown). The emitting agents at the visible wavelengths are primarily TiO and VO, similar to the dayside, and those fluxes trace the strong heated thermal layers near the terminators (right panel of Figure \ref{fig.tps}). The nightside emission features of TiO and VO should be readily detectable in the short wavelength range by JWST, and the near-limb heat transport could be tested. 

Thermal spectra of models with stellar temperatures of 5500 K and 6000 K have qualitatively identical features except that they generally have higher dayside emission and lower nightside emission. This is due to the poor day-night heat transport when the rotation rate increases, which is more visible in the phase curves discussed below. 

\subsubsection{Cases with a  strong drag}
Similar to previous GCM parametric surveys of \cite{perezbecker2013,komacek2016,komacek2017}, we performed the same set of models as those partly shown in Figures \ref{fig.temperaturemap24} and \ref{fig.temperaturemap40} but with a strong drag of $\tdrag=10^4$ s. Including the drag is motivated by the presence of possible magnetohydrodynamic dissipation (e.g., \citealp{perna2010,rogers2014komacek,beltz2022}) or mixing by small-scale turbulence that cannot be resolved in global models (e.g., \citealp{li2010}). The simple drag scheme crudely mimics the dissipation of kinetic energy of the flow over a global scale in both the zonal and meridional directions. The strong drag typically reduces the wind speed, damps out the superrotating jet and the global-scale Matsuno-Gill wave pattern, and increases the day-to-night temperature variations. The horizontal pattern is instead mostly characterized by the day-to-night flow, except for some  rapidly rotating cases in which the magnitude of the Coriolis force is comparable to the drag force and  global wave patterns still appear \citep{showman2013}. Therefore, compared to the strong-drag cases in  \cite{tan2019uhj} in which the primary model sets assumed a fixed 2.43-day rotation period, the relatively rapid rotating, strong-drag cases in this study exhibit more obvious wave patterns. For more discussion of the circulation and temperature maps, readers are referred to \cite{tan2019uhj} for results relevant to ultra-hot Jupiters and we do not show them here. 

\begin{figure*}      
\centering
\includegraphics[width=2.\columnwidth]{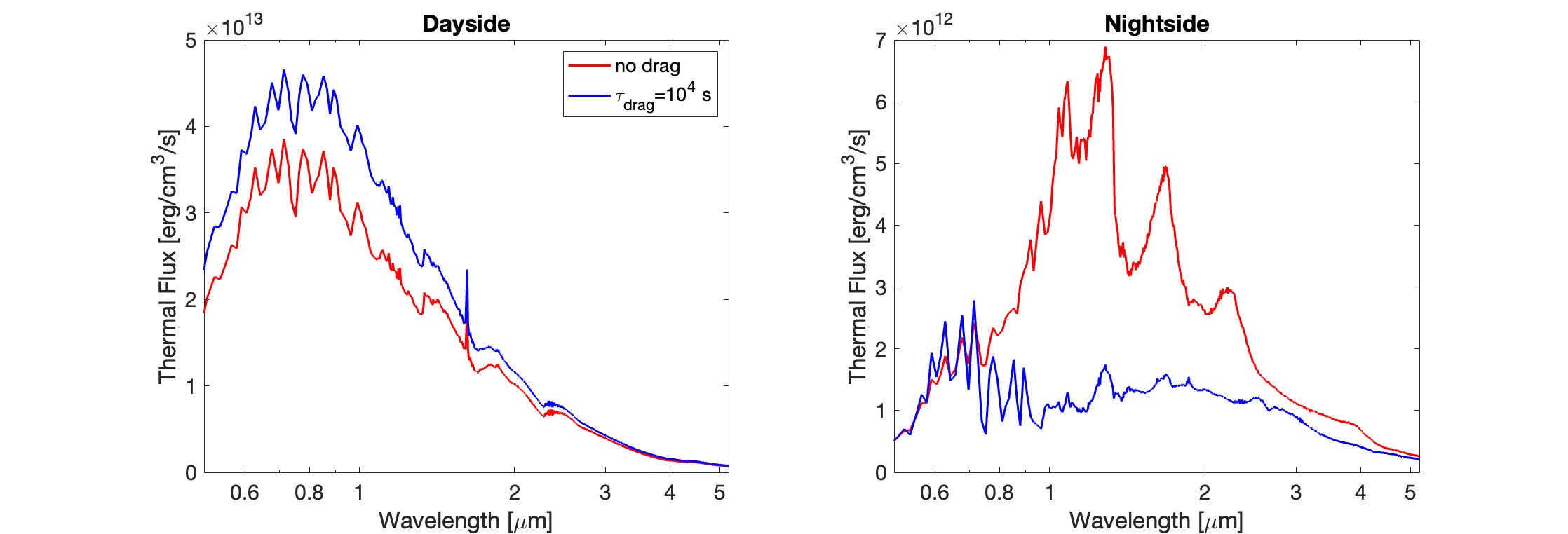}
\caption{Modeled thermal emission spectra from the dayside (left panel) and from the nightside (right panel). Red lines are from the model of a planetary $\teq=2600$ K, a stellar temperature of 6500 K and without drag; blue lines are from a similar model except for a strong drag with a timescale  $\tdrag=10^4$ s.}
\label{fig.spectra_dragcomparison}
\end{figure*}

On the dayside, as far as the vertical temperature structures, heating profiles, and the resulting thermal spectra, those of the strong-drag models are qualitatively similar to those of drag-free models (for temperature structures, see the left panel of Figure \ref{fig.tps} and the upper left panel of Figure \ref{fig.noTiOVO}). The primary quantitative difference is that the strong-drag models have greater dayside fluxes and reduced nightside fluxes due to their poorer day-to-night heat transport;  the spectral features are almost identical otherwise. This can be seen in the comparison in the left panel of Figure \ref{fig.spectra_dragcomparison} for models with or without drag, both of which have $\teq=2600$ K and a stellar temperature of 6500 K. 

On the nightside, the total spectral energy flux is significantly reduced in the strong-drag models, as expected from the poorer heat transport. However, in the case shown in the right panel of Figure \ref{fig.spectra_dragcomparison}, the nightside comparison becomes much more subtle than the dayside comparison, showing much more muted absorption features in the strong-drag case in near-IR and IR wavelengths; and some regions seem to be blended with a weak emission feature within a broader absorption feature, for example, around the water 1.4 $\mu$m window; as well as comparable or even greater flux at short wavelengths $< 0.7~\mu$m. The nightside ``nonstandard" near-IR and IR spectral properties of the strong-drag model come from two distinct contributing areas, 1) near-limb regions where there are strong thermal inversions and fluxes are generally emitted at low pressures between 1 to 10 mbar, and 2) the broad area around the anti-substellar point where there is no inversion and the photospheric pressures are higher (pressure slightly less than 1 bar). Whereas in the drag-free model, the nightside fluxes at near-IR and IR are primarily from the broad anti-substellar areas (with much less flux contributed from the limb, low-pressure regions) and therefore reflect the non-inverted temperature structures of that region. In the visible wavelengths, both models probe the low-pressure limb regions, and so they show similar flux.  The above findings were from constructing the 3D contribution functions by perturbing the 3D temperature field and rerunning the radiative transfer post-processing every time a local temperature is perturbed. Our results highlight that the nightside spectrum of some UHJs may show exotic features depending on the detailed circulation pattern, and JWST should be able to detect these possible features \citep{mikal-evans2023}.

\subsubsection{Cases without TiO, VO, and Fe}

\begin{figure*}      
\centering
\includegraphics[width=2.\columnwidth]{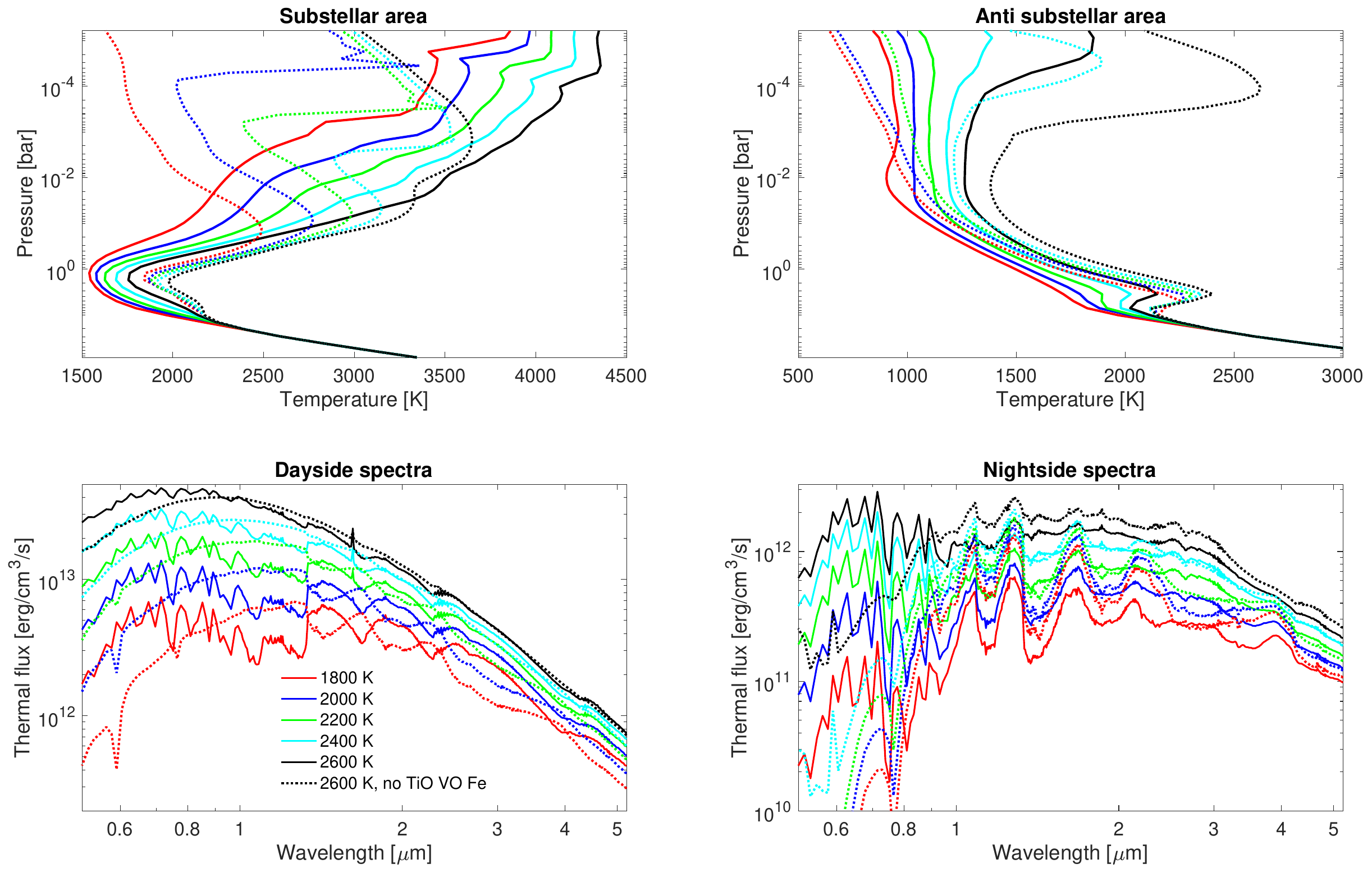}
\caption{Temperature-pressure profiles on the dayside (upper left panel) and nightside (upper right panel) for cases either including (solid lines) or excluding TiO, VO, and Fe (dotted lines). Corresponding thermal  spectra on the dayside and nightside are shown in the bottom panels. All GCMs have a drag timescale of $\tdrag=10^4$ s and a stellar temperature of 6500 K.}
\label{fig.noTiOVO}
\end{figure*} 

Optical absorbers such as TiO and VO have long been known to significantly affect the atmospheric structure, chemistry, and dynamics of hot Jupiters (e.g., \citealp{hubeny2003,fortney2008unified,showman2009}). The search for these species using emission and transmission spectroscopy so far has not yielded a common existence of TiO and VO in the atmospheres of hot Jupiters, potentially related to the cold trap of these species \citep{spiegel2009,parmentier2013}. More recently, additional metal species along with wind velocities have been found to exist in the atmospheres of some ultra-hot Jupiters via high-resolution spectroscopy (e.g., \citealp{yan2019ionized,gibson2020,Hoeijmakers:2020aa,PaiAsnodkar2022}) and should contribute to significant heating in the upper atmospheres (e.g., \citealp{lothringer2018}). Fe is one of the most important species among them, which is also included in our main suite of GCMs. Similar to TiO and VO, Fe is also condensible in atmospheres of hot Jupiters \citep{helling2019, gao2021} and might be subjected to the cold trap process. Notably, observations of the ultra-hot Jupiter WASP76b indicate the partial rain-out of Fe \citep{Ehrenreich:2020aa}, but the transit asymmetry in Fe absorption could instead be due to clouds or scale height variations \citep{Savel:2021aa,Wardenier:2021td}. 

Here we explore the effects of excluding the strong optical absorbers on the thermal structure, heat transport, and resulting thermal spectra of ultra-hot Jupiters. Several species are excluded (see Table \ref{table:params}), most importantly TiO, VO and Fe. For this test, we include a strong drag with $\tdrag=10^4$ s to accelerate the convergence of the models and study a limiting case with reduced day-night heat transport. Figure \ref{fig.noTiOVO} summarizes our findings. The dayside and nightside temperature profiles are shown in the top row. Near the infrared photosphere on the dayside, there is no strong thermal inversion in cases without strong optical absorbers except in the hottest case in which water cooling is weak. Correspondingly, the dayside thermal emission spectra of these cases (dotted lines in the lower left panel) display either a near blackbody in the hotter cases or absorption features in the molecular bands in cooler cases, in contrast to those of the canonical suite of models that show strong emission features (solid lines). Because of the weak  absorption in the upper atmosphere, more stellar flux can be deposited in more depths, leading to  hotter infrared photospheres in models without TiO, VO and Fe due to the stronger infrared greenhouse \citep{guillot2010}. For more in-depth discussion of the impact of metallic absorbers on the temperature structure and emission spectra of ultra-hot Jupiters, see \cite{parmentier2018} and \cite{lothringer2018}.

The effect on the nightside is also profound. The nightside thermal photospheres of models without TiO, VO and Fe are also hotter due to more heat being transported from their hotter dayside thermal photosphere, which is consistent with the finding in the pseudo-2D framework of \cite{roth2021}. The nightside spectra shown in the lower right panel of Figure \ref{fig.noTiOVO}, as well as those shown in Figures \ref{fig.spectra-dragfree} and \ref{fig.spectra_dragcomparison}, demonstrate that even on the nightside, strong TiO and VO spectral features can be readily detectable in models with TiO and VO included.   Interestingly, cases with $\teq=2400, 2600$ K have obviously hotter upper atmospheres than those including strong optical absorbers. The optical absorbers of TiO, VO and Fe act as cooling agents on the nightside which would in principle radiate away heat transport if they were included, but in models without those absorbers the upper-layer temperature remains high at low pressures due to the efficient heat transport by hydrogen recombination and inefficient thermal cooling. 
The lack of thermal cooling by TiO and VO  can be seen in the lower right panel in which strong emission is absent in the optical wavelengths in cases without TiO, VO and Fe. We also post-process the GCM outputs without TiO, VO and Fe using k-coefficient tables that do include those species, and excessive  flux appears on the optical wavelength, indicating that if there were those species, the upper atmosphere should be effectively cooled down.  High-resolution spectroscopy observations might be able to measure the nightside-to-limb upper thermal structures and provide constraints on the cooling role of these species.

\subsubsection{Circulation influenced purely by different stellar spectra}
\label{ch.different_stars}

\begin{figure}      
\centering
\includegraphics[width=1.\columnwidth]{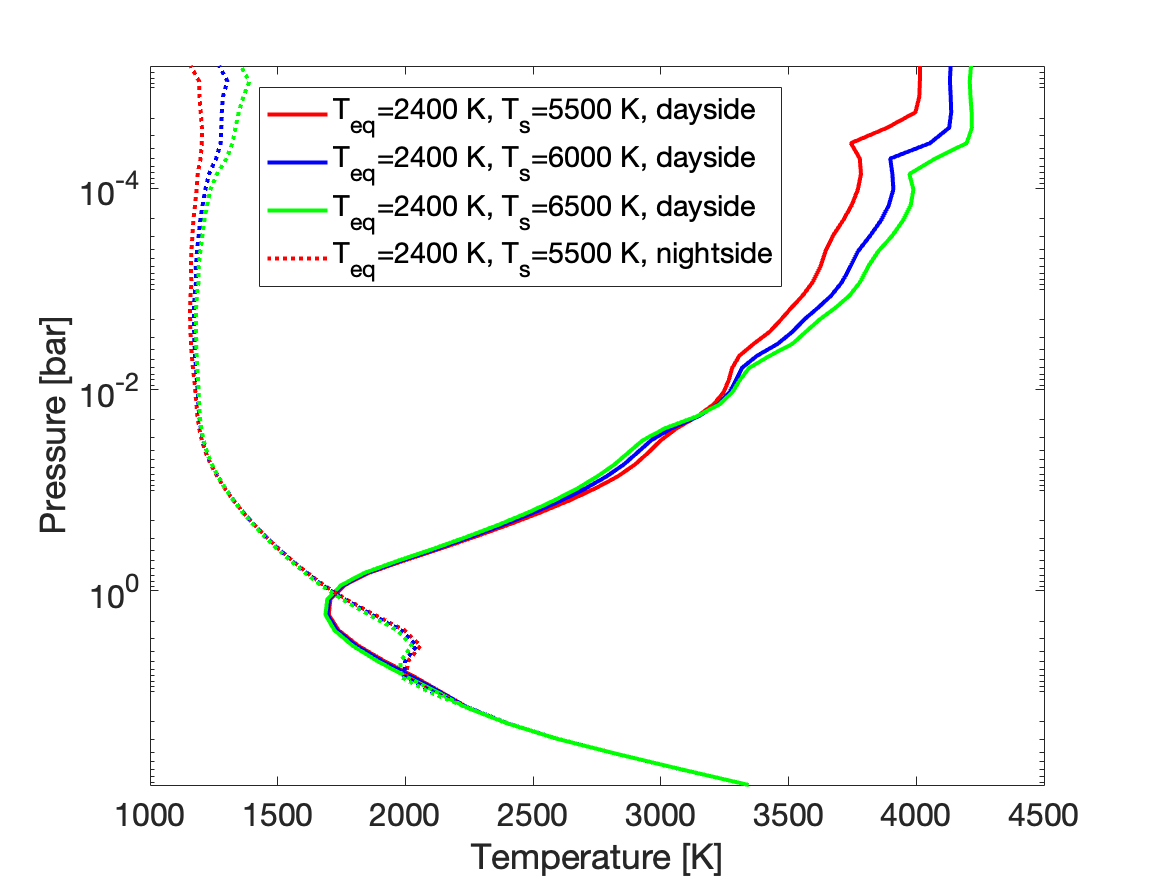}
\caption{ Temperature-pressure profiles on the dayside (solid lines) and nightside (dotted lines) for models with $\teq=2400$~K and all other parameters the same, including the same rotation period of 1.3 days, but only the stellar effective temperature differs from 5500 K to 6500 K. These profiles are from spatial averaging of the same manner as those shown in Figure \ref{fig.tps}. All GCMs have a drag timescale of $\tdrag=10^4$ s.}
\label{fig.tprfix}
\end{figure} 

The temperature structures of ultra-hot Jupiters could be sculpted by different stellar spectra as modeled by static 1D models of \cite{lothringer2019}, but their role in circulation is unclear. Results in our main suites of models are mixed with effect of varying rotation rates. To isolate the rotational effect,  we also performed a few additional models with $\teq=2400$~K and all other parameters the same, including the same rotation period of 1.3 days, but only the stellar effective temperature differs from 5500 K to 6500 K.  The temperature profiles are shown in Figure \ref{fig.tprfix}. On the dayside, the upper atmosphere is hotter and the infrared photosphere is cooler when the stellar temperature is higher. The nightside temperature structures show the same trend. This trend is in line with  \cite{lothringer2019}, and is because hotter stars have a higher fraction of the stellar energy in the shorter wavelengths that could be absorbed in the upper atmosphere by metals.   However,  the magnitude of the differences, especially in the thermal photosphere, is not significant in our work. This is not surprising  partly because the stellar temperature variations in this work are far less extreme than those explored in \cite{lothringer2019}  with stellar effective temperatures from 5700 K to 10500 K.   Dynamics might play a secondary role in regulating this difference by heat transport. Because of the relatively small effect on the thermal structure, the overall circulation is barely affected by the different stellar temperatures when other parameters are held fixed. Another difference is that models used in \cite{lothringer2019} reach shorter wavelengths and capture more UV fluxes than our GCM and \texttt{PICASO} whose shortest wavelength reaches about 0.26 $\mu$m. The UV absorption is important for the hottest stars in the UHJ samples and drives stronger thermal inversion at very low pressures, but this difference is expected to be minor here because of the less extreme stellar temperatures we considered here and that the low-resolution emission spectrum is generally less sensitive to structures at very low pressures. 

\subsection{Comparisons to observations: dayside brightness temperatures and phase-curve amplitudes}

\begin{figure*}      
\centering
\includegraphics[width=2.\columnwidth]{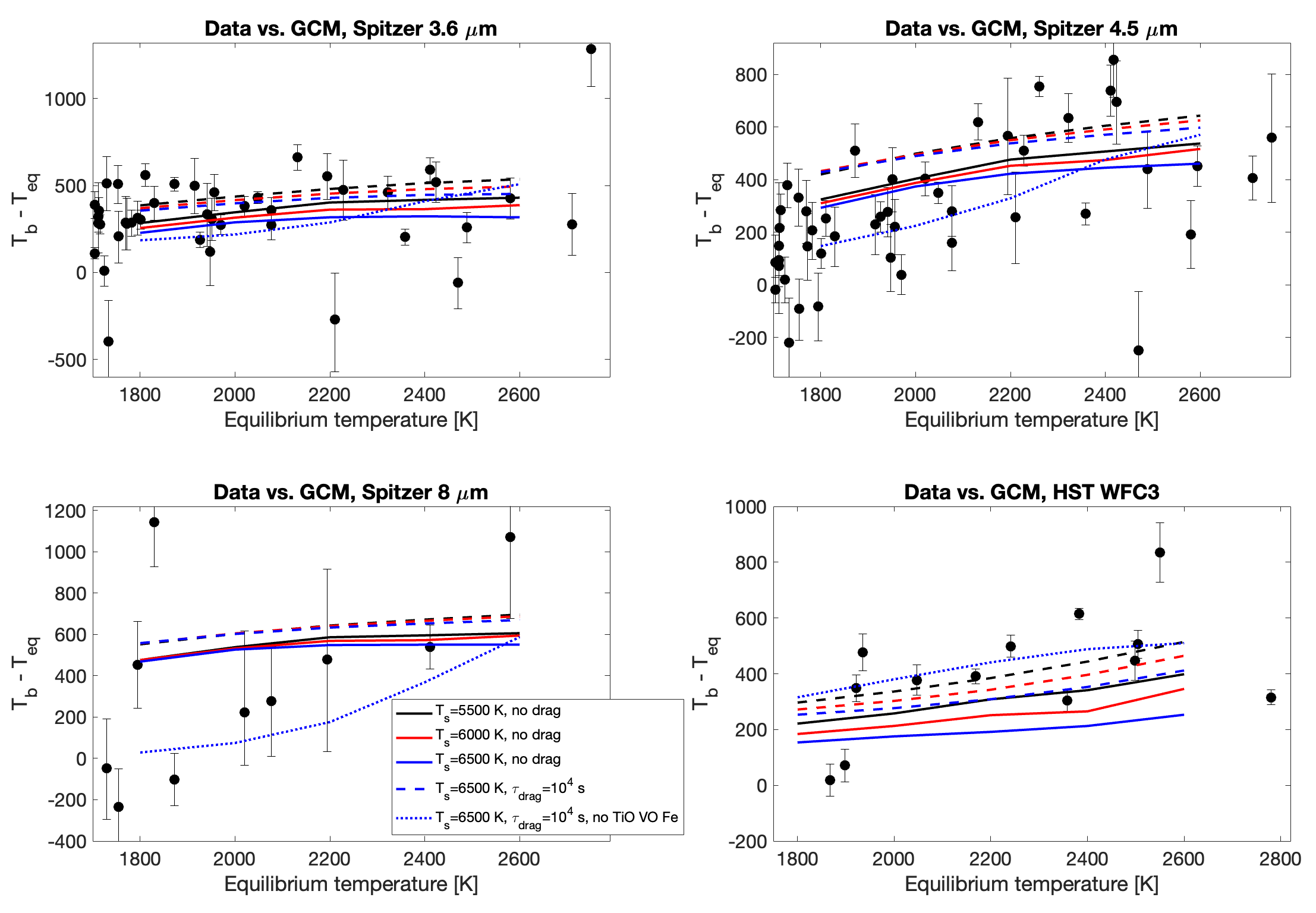}
\caption{Dayside brightness temperature minus the planetary equilibrium temperature as a function of planetary equilibrium temperature. Dots with error bars are {\it Spitzer} data presented in \protect\cite{deming2023} with different panels representing analysis at different {\it Spitzer} bandpasses, and  HST WFC3 data collected in \protect\cite{mansfield2021}. Lines are from different sets of models: black, red and blue lines are from models with stellar temperatures of 5500 K, 6000 K and 6500 K, respectively; solid lines are drag-free and dashed lines are with $\tdrag=10^4$ s; the dotted line is from models with $\tdrag=10^4$ s, the stellar temperature of 6500 K and without TiO, VO and Fe.}
\label{fig.compare_deming}
\end{figure*} 

\begin{figure*}      
\centering
\includegraphics[width=2.\columnwidth]{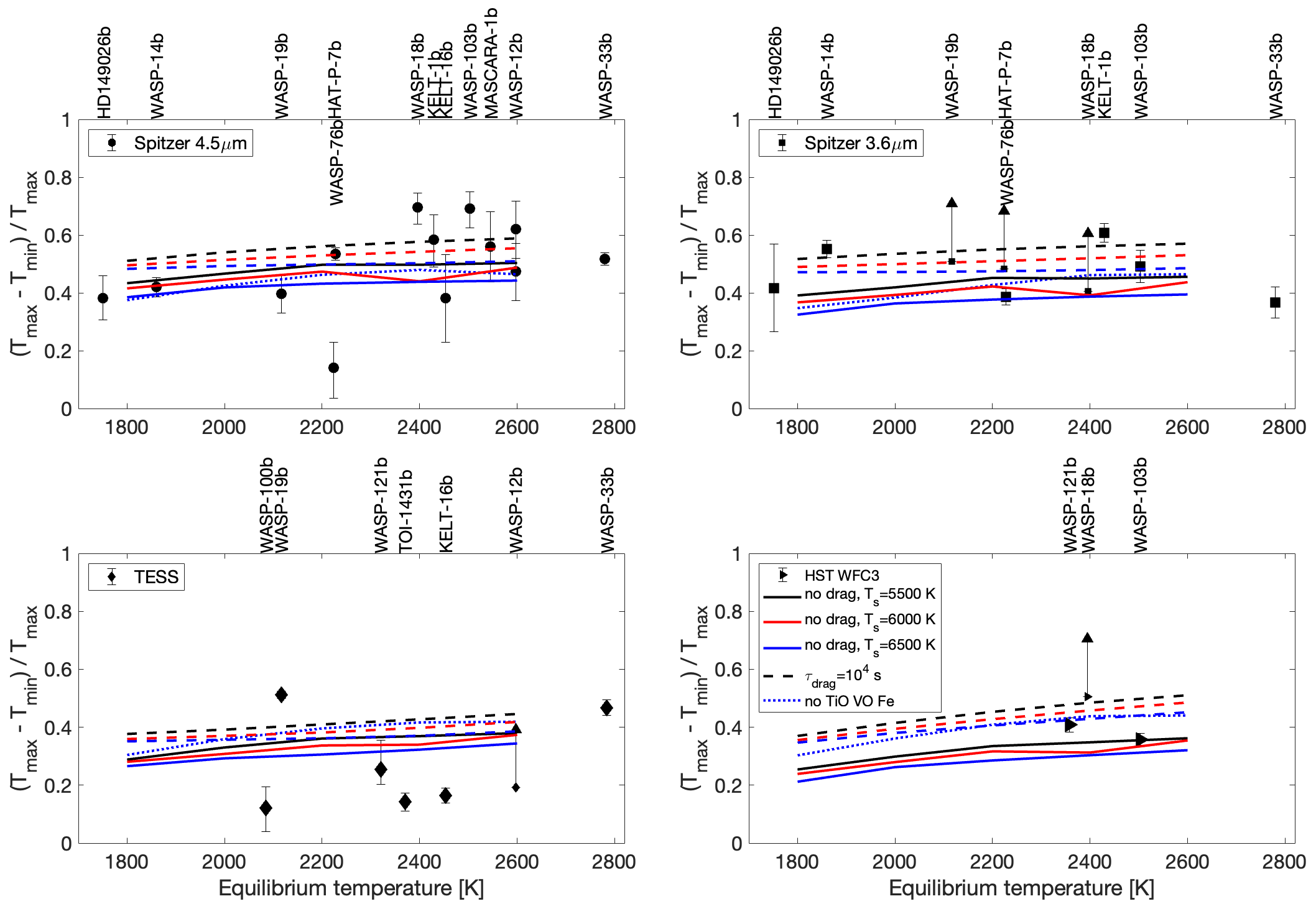}
\caption{Day-to-night brightness temperature differences as a function of equilibrium temperature. Lines are from  different sets of models that are described in Figure \ref{fig.compare_deming}, and dots with error bars are observations described below. They are separately  shown in  different instrument bandpasses as indicated above each panel. Data are updated from those summarized in \protect\cite{tan2019uhj} and \protect\cite{wong2021}. At {\it Spitzer} 4.5 $\mu$m, most data were from analysis by \protect\cite{bell2021}, WASP-76b is from \protect\cite{may2021}, HD 149026b is from \protect\cite{zhang2018spitzer}. At {\it Spitzer} 3.6 $\mu$m, data are from \protect\cite{wong2015,wong2016}, \protect\cite{maxted2013},  \protect\cite{cowan2012}, \protect\cite{beatty2019} \protect\cite{may2021} and \protect\cite{kreidberg2018}. HST WFC3 data are from \protect\cite{kreidberg2018,arcangeli2019,mikal2022}. 
TESS observations are obtained from \protect\cite{mancini2022}, \protect\cite{owens2021}, \protect\cite{addison2021}, \protect\cite{jansen2020}, \protect\cite{von2020}, \protect\cite{daylan2021}, and \protect\cite{eftekhar2022}. For WASP-121b TESS phase curve, \protect\cite{bourrier2020} and \protect\cite{eftekhar2022wasp121b} also show roughly consistent dayside and nightside brightness temperatures as \protect\cite{daylan2021}. }
\label{fig.daynighttemp}
\end{figure*}

\begin{figure*}      
\centering
\includegraphics[width=1.8\columnwidth]{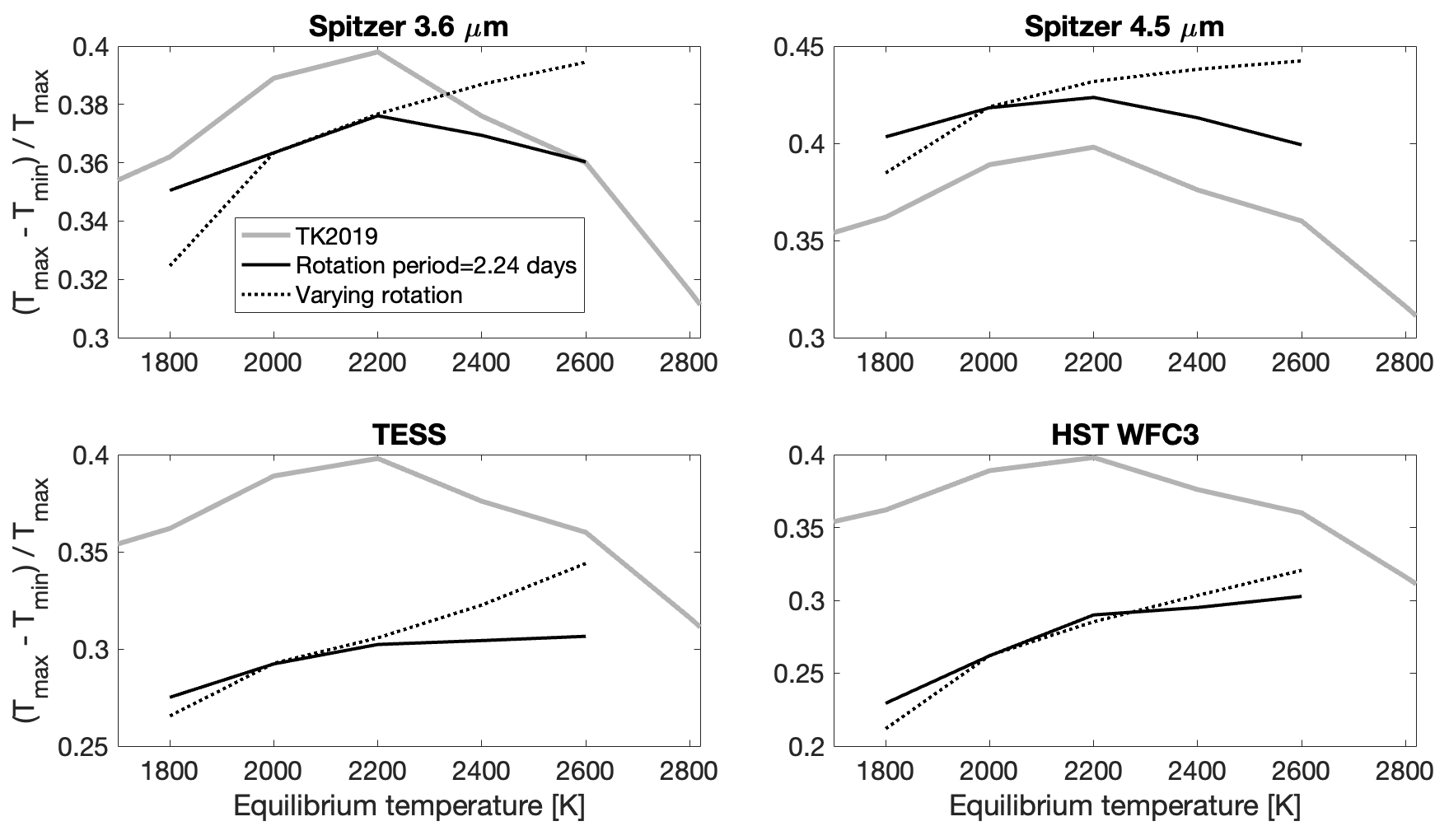}
\caption{Normalized day-to-night brightness temperature variations at bandpasses of different instruments. The black solid lines are from a set of simulations with a fixed rotation period of 2.24 days and different $\teq$ around a 6500 K star. The dotted lines are models with the same parameters except that the rotation periods are [3.07, 2.24, 1.68, 1.3, 1.06] days for $\teq=$ [1800, 2000, 2200, 2400, 2600] K.  The thick grey lines are normalized day-to-night temperature variations from a set of semi-grey GCMs with ${\rm H_2}$-H conversion and $\tdrag=10^7$ s in \protect\cite{tan2019uhj} (TK2019). TK2019 models assumed a fixed rotation period of 2.43 days, and these results were derived from the bolometric fluxes and do not distinguish between different instrument bandpasses (grey lines are identical in all panels). The TK2019 results are included for comparisons between GCMs with realistic non-grey and idealized semi-grey radiative transfer schemes.  }
\label{fig.daynight_model}
\end{figure*}

\begin{figure}      
\centering
\includegraphics[width=1\columnwidth]{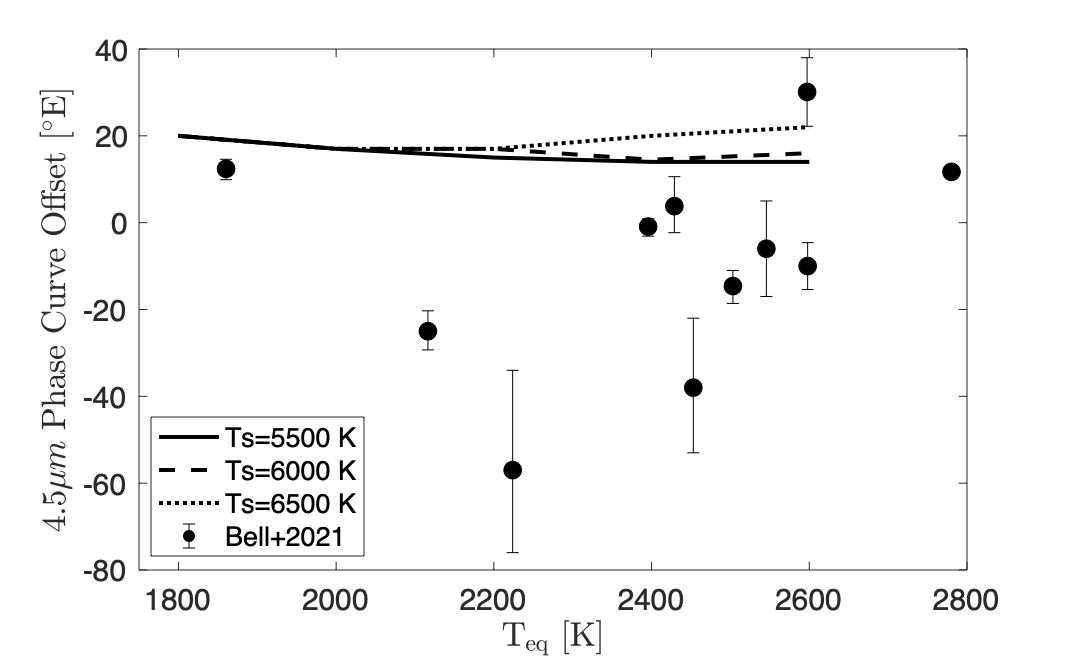}
\caption{Phase offsets of drag-free models with different equilibrium temperatures for Spitzer 4.5$\mu$m and comparisons to those presented in the uniform analysis of \protect\cite{bell2021}. Different lines   represent different stellar effective temperatures.
}
\label{fig.phaseoffset}
\end{figure} 

Using post-processed phase-resolved emission spectra by \texttt{PICASO}, we are able to directly compare the  model-predicted dayside emission and phase-curve amplitudes with observed values over the population of ultra-hot Jupiters.

We first compare the dayside photometric brightness temperature over the {\it Spitzer } 3.6, 4.5 and 8 $\mu$m bands as well as the HST WFC3 band in Figure \ref{fig.compare_deming}. Dots with uncertainties in the three {\it Spitzer} bands are from the uniform reanalysis of \cite{deming2023}, which is based on a similar analysis of \cite{garhart2020} but with more updated targets. Data in the HST WFC3 band are summarized in \cite{mansfield2021}. The observational data sets extend to lower effective temperatures but here we only show those above about $\teq=1700$ K for comparisons with our model grids. Lines in the plots represent model outputs, including cases with three stellar temperatures,  drag-free, and $\tdrag=10^4$ s, as well as one set that excludes TiO, VO and Fe. The vertical axis is the brightness temperature at that bandpass minus the $\teq$ of the planets.

Regarding just the model-predicted dayside brightness temperatures,  there is a general trend of rising brightness temperature with increasing $\teq$ at all wavelengths and all model assumptions.  Models with a strong drag of $\tdrag=10^4$ s have overall higher brightness temperatures than drag-free ones due to the inefficient day-to-night heat transport. At a given drag setup, models with a cooler stellar temperature exhibit higher dayside brightness temperature because these planets rotate faster which inhibits day-to-night heat transport more efficiently \citep{tan2020wdbd}.  In Section \ref{ch.different_stars}, we have discussed that results are quantitatively similar among models with different stellar types but all other parameters are the same. Here we demonstrate that with a reasonable stellar type range, the effects of rotation dominate over the effects of the different stellar spectra in the circulation, emission temperature, and day-night flux differences. 
The dayside brightness data (dots in Figure \ref{fig.compare_deming}) show a general agreement to models with a wide range of parameters in terms of the trends with increasing equilibrium temperature; however, the  scatter of data is much wider than the model predictions, indicating a much wider intrinsic parameter scattering among the hot Jupiter population or that our model is missing some other key processes such as the formation of clouds and their effect on the heat transport (e.g., \citealp{roman2021, parmentier2021}). 

The dayside brightness temperature at a certain wavelength of hot Jupiters is affected by both the radiative and dynamical processes, and the question is which process is more important at a certain wavelength. Figure \ref{fig.dayside_model} shows the same model predictions as Figure \ref{fig.compare_deming} but without observational data for better visualization. We compare the dayside brightness temperature variations caused by different orbital and drag parameters (but with the same atmospheric compositions) to those caused by different atmospheric compositions (but the same orbital and drag parameters), and find that at relatively shorter wavelengths, for example, the TESS and HST WFC3 bandpasses, day-night heat transport is as important as the radiative processes. But towards longer wavelengths from {\it Spitzer} 3.6 to 8 $\mu m$, the radiative process becomes more and more important than the day-night dynamical heat transport. 

Next, we turn our attention to the day-to-night variations of brightness temperatures.
Figure \ref{fig.daynighttemp} shows the normalized day-to-night brightness temperature variation for all models, together are observed values for ultra-hot Jupiters at various bandpasses. Lines are modeled outputs and are labeled the same as those in Figure \ref{fig.compare_deming}. The peak and trough thermal flux over individual bands are converted to brightness temperatures, which are then normalized to get those values. The day-night brightness temperature variations are less sensitive to whether there are strong visible absorbers, in contrast to the dayside brightness temperature shown in Figure \ref{fig.compare_deming}. 
Given a fixed host star, the fractional day-night brightness temperature variations at different wavelengths all exhibit either a nearly constant or slight increase with increasing equilibrium temperature, which is in contrast to analytic analysis of \cite{komacek2018rnaas} and GCMs with simplified radiative transfer \citep{tan2019uhj}, in which the fractional variations first increase but then decrease with increasing equilibrium temperature. 
The previous work assumed a fixed rotation period of slightly more than 2 days for models with different equilibrium temperatures. This difference is due to the increased rotation rate with increasing equilibrium temperature around a given star in this work, and the increasing rotation rate tends to reduce the day-night heat transport efficiency as discussed above. 

We can notice a few obvious points from the data-model comparisons. First, the range of our model predictions is still less than the observed values of the day-night variations especially in the {\it Spitzer} 4.5 $\mu$m band and the TESS band, although our models have covered a wide range of drag setup, rotation periods and stellar properties. Second, in the TESS band, there seems to be a systematic overestimate of the day-night temperature variation compared to the observed values. This is a tougher issue than increasing the day-night temperature contrasts because the latter can be potentially achieved in several self-consistent ways such as nightside clouds and strong magnetohydrodynamic drag. In contrast, some of our GCMs are  nearly drag-free (apart from numerical dissipation which is necessary to maintain numerical stability) and include hydrogen dissociation and recombination which are already prompted for the most efficient day-night heat transport. One remote possibility is that the planets have enormous internal heat flux (even much larger than those estimated by \citealp{Thorngren:2019aa}) that is comparable to the incoming stellar flux due to unknown processes, but it is implausible to argue that the TESS observed population all have that amount of internal heat flux. The scenario of large internal heat flux is also inconsistent with their present-day radius. The systematic overestimation of the modeled day-night temperature differences in the TESS band motivates us to look for alternative solutions to enhance heat transport in the Appendix \ref{sec:metallicity}.    Third, for objects with  phase-curve observations over multiple wavelengths, no single model setup can simultaneously explain all the observations. One notable example is WASP-103b, for which the {\it Spitzer} 4.5 $\mu$m phase-curve amplitude is much larger than our grid; the {\it Spitzer} 3.6 $\mu$m one lies in the $\tdrag=10^4$ s branch of models; lastly the HST WFC3 one is in the drag-free branch of models. Others include WASP-18b, WASP-14b, WASP-19b, WASP-121b, and WASP-33b. 

We conducted an additional set of simulations with a fixed rotation period of 2.24 days and different $\teq$ around a 6500 K star. Results are shown as the black solid lines in Figure \ref{fig.daynight_model}. Dashed lines are included for comparisons and they are from models with the same parameters except that the rotation periods are [3.07, 2.24, 1.68, 1.3, 1.06] days for $\teq=$ [1800, 2000, 2200, 2400, 2600] K as described in Section \ref{sec:basic}.  The comparisons between the two sets of models immediately reveal the importance of rotation rate in controlling the day-night heat transport. Interestingly, at {\it Spitzer} 3.6 and 4.5 $\mu$m bands, the fixed-rotation-rate models show a fractional variation turning point at about 2200 K which is in good agreement with previous work of \cite{komacek2018rnaas} and \cite{tan2019uhj} that used more idealized models; however, the fractional variations at TESS and HST WFC3 bandpasses keep increasing with increasing temperature, although the trend slows down after $\teq=2200$ K, unlike the IR bands and those predicted in \cite{komacek2018rnaas} and \cite{tan2019uhj}.  Finally, results from a set of semi-grey models with ${\rm H_2}$-H conversion, $\tdrag=10^7$ s and a fixed rotation period of 2.43 days in \cite{tan2019uhj} are included as thick grey lines in Figure \ref{fig.daynight_model} for comparisons between realistic and simplified models. The qualitative trend of the semi-grey curve agrees with the non-grey results in the Spitzer bands when rotation periods are held fixed (grey lines vs the black solid lines) to some extent.  However, the non-grey results exhibit both qualitative and quantitative differences from the semi-grey results through the influences of the thermal structure by non-grey heating/cooling rates and also through non-grey opacity (meaning that different wavelengths probe different atmospheric layers). For example, the trends are qualitatively different in the WFC3 and TESS bands with significant offsets between different modeling frameworks, and the absolute values differ obviously in the {\it Spitzer } bands.   These again highlight the wavelength-dependent nature of the thermal phase-curve amplitudes for the hot Jupiter population \citep{parmentier2021}. 

Lastly, we compare the model-predicted phase curve offsets with those from the reanalysis of {\it Spitzer } 4.5 $\mu$m phase curves \citep{bell2021} in Figure \ref{fig.phaseoffset}. Only models with no drag are included; the phase curve offsets of models with $\tdrag=10^4$ s are generally very small (but are still positive) due to the suppression of jets and waves. The modeled offsets are in general smaller than $\sim22^{\circ}$ and larger than $\sim15^{\circ}$ (positive values indicate the eastward displacements of the hot regions relative to the sub-stellar point). This narrow variation is in large contrast to the observed values that range from almost $-60^{\circ}$ to $30^{\circ}$. In addition, many observed offsets are negative which is not predicted by many GCMs of hot Jupiters (see a review of \citealp{showman2020}). The factors affecting the thermal phase-curve offsets include magnetohydrodynamics (e.g., \citealp{rogers2017}) and clouds (e.g., \citealp{parmentier2021,roman2021}), for which we do not include in our models.  Given the incompleted physics in our models and the potential sensitivity to data analyzing methods, we may say that the drag-free and cloud-free models are likely insufficient to explain the observed phase-curve offsets. Lastly, the phase offsets can be sensitive to different data processing methods, for example, some data collected in \cite{tan2019uhj} and presented in \cite{may2022} are different from those reported in \cite{bell2021} for the same objects. Even in \cite{bell2021}, for many planets with negative phase offsets, the phase offsets could be either positive or negative depending on the data reduction methods (see their Figure 5). 
JWST will help to break the current conflicting phase-curve observations of some targets (e.g, WASP-121b) with exquisite precision and (simultaneously) large wavelength coverage for hot Jupiter atmospheres \citep{mikal-evans2023}.

\subsection{Variability}

Searching for time variability of hot Jupiters' atmospheres is a timely topic, helping to constrain the dynamical mechanisms affecting the meteorology of these objects (e.g., \citealp{armstrong2016variability, jackson2019variability, wong2022variability}). \cite{komacek2019temporal} investigated the variability of  hot Jupiters using a suite of GCMs and found that about 2\% variability can be expected but the origin of the variability was mostly confined near the equator. \cite{tan2019uhj} and \cite{tan2020wdbd} similarly found variability in GCMs with short rotation periods, and the mechanism is similar to those shown here. In \cite{tan2020wdbd}, the zonal-mean meridional potential vorticity is shown to have reversed signs in the vertical direction in relatively rapid rotators of tidally locked hot Jupiters and  the baroclinic instability is the most likely mechanism here. 
Some GCM studies with ultra-high spatial  resolution and low numerical dissipation showed that significant evolving storms may occur even under numerical setups for canonical hot Jupiters (e.g., \citealp{skinner2022}), in contrast to many other GCMs \citep{showman2020}. 

\begin{figure*}      
\centering
\includegraphics[width=2.\columnwidth]{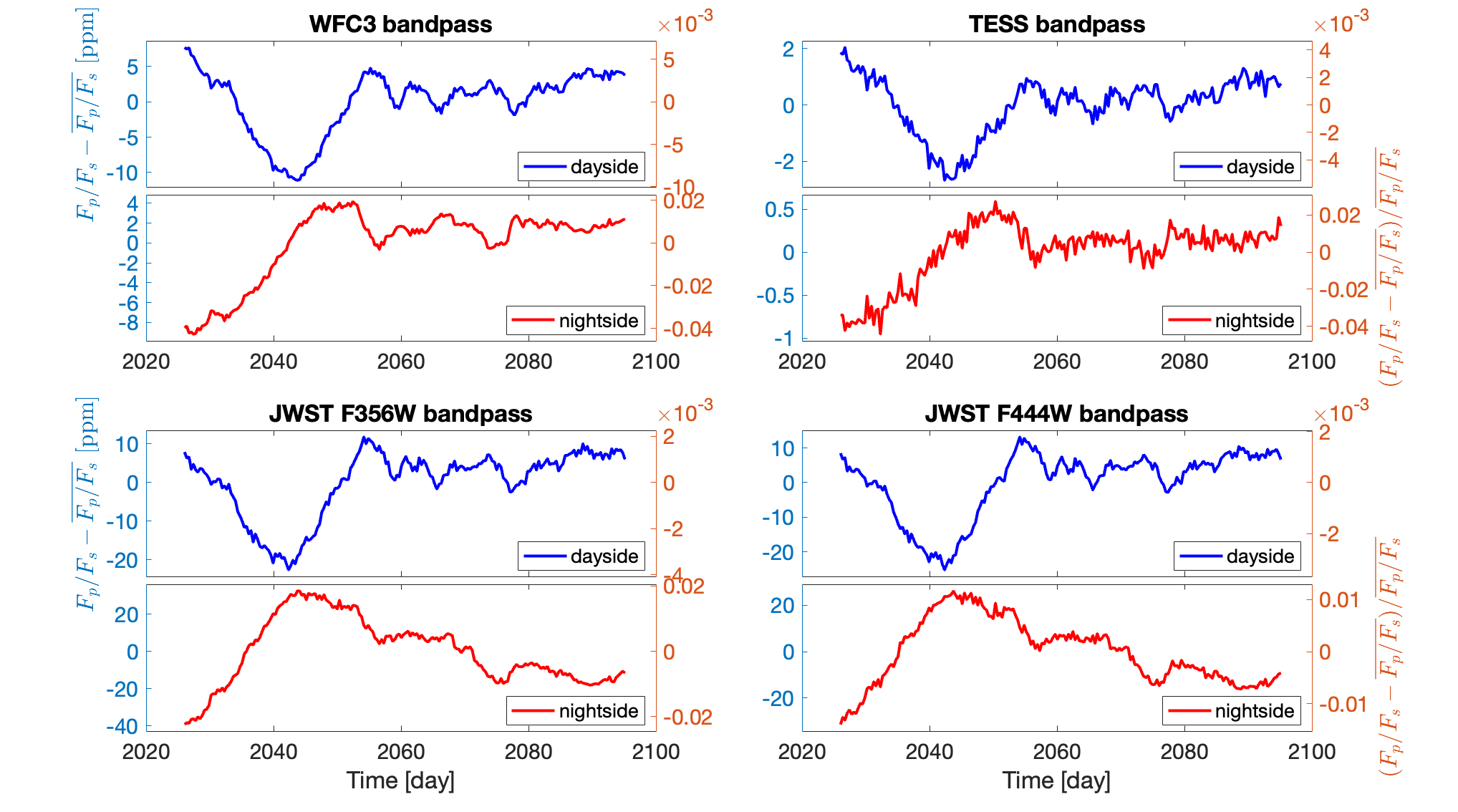}
\caption{Variability of the planet-to-star emission ratio relative to the time-mean  during the output time from the model with an equilibrium temperature of 2000 K around the 5500 K star. Blue lines are from the dayside spectra and red lines are from the nightside spectra. The flux ratios through different bandpasses are shown in different panels. The left vertical axis of each panel is variability measured by part per million, and the right vertical axis represents the fractional variation relative to the time mean. }
\label{fig.variability}
\end{figure*} 

Our drag-free simulations show variability in their dayside and nightside temperatures. Figure \ref{fig.variability} shows the dayside and nightside emission as a function of time from the case with $T_{\rm eq}=2000$ K around the 5500 K star. Long-term variation modulations with timescales of tens of days dominate the variability, and variability with much lower amplitudes also presents over the orbital timescale. The amplitude of the hemispherically integral emission can reach over 60 ppm at wavelengths near 3.6 $\mu$m and over 40 ppm at wavelengths near 4.4 $\mu$m; but the variability amplitudes are relatively small at visible and near-IR wavelengths where they probe higher-temperature layers and thus the fractional flux perturbations caused by certain temperature variations are smaller.   A visual inspection of the temporal evolution of the global temperature maps of these simulations suggests that the long-term variation is driven mostly by off-equatorial eddies resulting from atmospheric instabilities on the dayside and then subsequently migrating longitudinally to the nightside and dissipated. Although the fractional variability is similar at different bandpasses, their amplitudes at planet-to-star flux ratio differ simply because of the different basic planet-to-star ratios at different wavelengths. Therefore, the emission variability on the dayside or nightside would be best tested in infrared wavelengths compared to optical or near-IR wavelengths. JWST has yielded high-precision secondary eclipse observations of (ultra) hot Jupiters' atmospheres \cite{brightstar2023}; the currently predicted variability amplitudes of over 60 ppm at about 3.6 $\mu$m wavelengths on the nightside and about 30 ppm on the dayside should be within the limit of JWST single-eclipse precision. For optimal objects and at infrared wavelengths, JWST should be able to put constraints on the meteorological variability mechanisms proposed on hot Jupiters.

\section{Summary and prospects}
\label{sec:summary}
\subsection{Summary}
The day-night temperature variations inferred from phase-curve observations of ultra-hot Jupiters provide rich information about the mechanisms controlling the atmospheric circulation of these planets, including radiative heating and cooling, dynamics of jets and waves, heat transport aided by hydrogen dissociation and recombination, cloud formation and their radiative feedback, and magnetohydrodynamics. In this study, we have upgraded our GCM for ultra-hot Jupiters \citep{tan2019uhj,komacek2022} to couple the dynamical core to a non-grey correlated-k radiative transfer scheme. This allows us to capture the more realistic atmospheric thermal structure and radiative heating and cooling rates. Using this model, we explore a wide range of parameter space including different planetary equilibrium temperatures from $\teq=$1800 to 2600 K, different tidally synchronized planetary rotation periods associated with combinations of planetary $\teq$ and varying stellar effective temperatures from 5500 K to 6500 K, and different frictional drag setup in Section \ref{sec:results}. 
To summarize our results in Section \ref{sec:results}, we find that:
\begin{itemize}
    \item Assuming a fixed solar composition of the atmosphere, the atmospheric dynamics of ultra-hot Jupiters depends sensitively on equilibrium temperature and host star type, and thus rotation rate. For a fixed equilibrium temperature, hot Jupiters orbiting earlier-type host stars have smaller day-night temperature contrasts due to their longer rotation periods assuming spin-synchronization. We also find that ultra-hot Jupiters with larger equilibrium temperatures orbiting a fixed host star type have greater day-night temperature contrasts and thus a faster speed of the equatorial jet. In agreement with our previous double-grey models, we find that hotter ultra-hot Jupiters will have smaller hot spot offsets. We also similarly find that strong drag further increases day-night temperature differences and reduces hot spot offsets.
    \item Our non-grey GCMs of ultra-hot Jupiters across a range of $\teq$ ubiquitously have strong dayside thermal inversions. In cases without strong drag, these thermal inversions persist to both the east and west limbs, with west limbs typically having higher-altitude thermal inversions than the east limbs in these cloud-free simulations. The dayside inversions are driven by shortwave absorption, which is largely balanced by cooling from H$_2$ dissociation on the dayside -- conversely, H$_2$ recombination is the dominant balance of radiative cooling on the nightside. By post-processing these models to predict emergent spectra, we confirm with 3D non-grey GCMs the expectation from 1D models that ultra-hot Jupiters will have dayside emission features with emission strength decreases with increasing equilibrium temperature.  Our cloud-free, drag-free simulations also generally display nightside absorption features, as nightside thermal inversions are limited to the upper atmosphere in our hottest cases. The dayside spectral properties are not sensitive to our model drag strength, but the nightside spectral shape could show significant differences between the drag-free and strong-drag cases. 
    \item Our GCM simulations of ultra-hot Jupiters that include the thermodynamic effect of hydrogen dissociation and recombination coupled to a non-grey radiative transfer scheme predict a weaker temperature-dependence of dayside temperatures and day-night temperature contrasts than previous non-grey models. We find that given the orders of magnitude uncertainty in $\tau_\mathrm{drag}$, our non-grey GCMs can broadly reproduce the observed dayside temperatures and day-night temperature contrasts from Spitzer and HST. In general, we require strong drag to explain observed ultra-hot Jupiter day-night temperature contrasts at Spitzer Channel 1 and in the HST/WFC3 bandpass. In tandem, the phase curve offsets of many ultra-hot Jupiters are near zero or even negative (i.e., implying that peak flux occurs after secondary eclipse). This may imply that MHD effects shape the observable properties of ultra-hot Jupiters. Weather TiO and VO are included has a strong impact on the dayside brightness temperatures especially for those at longer IR wavelengths. Besides the dayside spectra, our modeled nightside spectra also displace emission features of TiO and VO at visible wavelengths and they originate from relatively low-pressure layers near the limb regions, where strong thermal inversions occur and penetrate towards the nightside due to the day-to-night heat transport.

\end{itemize}

In addition, for one particular test case at a given planetary $\teq$, we explore the effects of varying atmospheric metallicity and carbon-to-oxygen ratio in Section \ref{sec:metallicity}, finding that:
\begin{itemize}
    \item At a given set of system parameters appropriate for TOI 1431b, we investigated the effects of varying metallicity from 0.1 to 3 of the solar value, and C/O ratio from 0.5 to 2.5 of the solar value, on the circulation and day-night temperature variation in the TESS band. Reducing metallicity reduces the day-night temperature variation and this trend is robust over different C/O ratios. At a given metallicity, varying C/O ratio changes the day-night temperature variation but the trend is less clear. The basic idea is the efficiency of radiative cooling decreases with decreasing metallicity and in some cases also the C/O ratio. Despite a large variation over this parameter space, the lowest day-night temperature variation in the TESS band in our models is still significantly greater than that observed. Further development of physical processes in the GCM and/or additional observations should be conducted to determine the source of this tension.
\end{itemize}

\subsection{Prospects and future work}
Our results serve as a baseline study for comparisons with those including clouds and varying parameters such as atmospheric composition and gravity. Clouds appear to be ubiquitous in atmospheres of close-in exoplanets. Their effects on transmission, reflection, and thermal emission have been explored in various contexts but their dynamic feedback in the general circulation of close-in exoplanets is comparatively less understood. If clouds show a significant non-grey effect on the thermal emission, phase curves at different wavelengths are essential to test the hypothesis that clouds are present \citep{parmentier2021}.  In parallel, mechanisms of cloud feedback in self-luminous planetary atmospheres have been shown to be important to drive variability and circulation of brown dwarfs and isolated giant planets \citep{tan2021bd2}. Despite that energy of the circulation in hot Jupiter's atmospheres is injected on the largest scale, we aim to explore whether mechanisms in self-luminous atmospheres are partly responsible for shaping the cloud structure and circulation. In \cite{komacek2022}, we started to incorporate the cloud radiative feedback in a hot Jupiter GCM using a semi-grey framework and showed that cloud patchiness could feed back on the dynamics and thermal structure. It is important to test the results in a more realistic (e.g., non-gray) modeling framework and over a wider planetary parameter space in future work. 


Another great challenge is the representation of MHD on hot close-in gas giant planets. Self-consistent MHD models are required to fully understand the dynamics of hot and ultra-hot Jupiters \citep{rogers2014komacek,rogers2017}. However, there have been relatively few non-ideal and self-consistent MHD models in the field to push the efforts further and confront observations, especially those of westward hot spot offsets. A more commonly seen approach is to parameterize the effects of (assumed) magnetic fields on the atmospheric flow in hydrodynamic models \citep{perna2010,Rauscher_2013,beltz2022}. 
In the future, adopting such an ``active'' kinematic magnetic drag scheme in our model could be an immediate next step to test our GCMs against observations. A long-term plan would be to move our current schemes of radiative transfer and thermodynamics treatments of hydrogen dissociation and recombination to self-consistent MHD models.

\section*{Acknowledgements}
X.T. and R.T.P.  acknowledge support from the European community through the ERC advanced grant EXOCONDENSE (PI: R.T. Pierrehumbert).  R. L. acknowledges support from the NASA ROSES XRP program with the grant 80NSSC22K0953, and from STScI grants JWST-AR-01977.007-A and JWST-AR-02232.008-A. The authors would like to acknowledge the use of the University of Oxford Advanced Research Computing (ARC) facility in carrying out this work  (\url{http://dx.doi.org/10.5281/zenodo.22558}). 
Part of this work was completed with resources provided by the University of Chicago Research Computing Center. 

\section*{Data availability}
The data underlying this article, including GCM outputs and post-processed phase-dependent spectra,  is publicly available in Zenodo \url{https://doi.org/10.5281/zenodo.10121933}. The 11-bin and 622-bin  k-coefficient tables used in this work can be  found in \cite{lupu111460,lupu6221460}.

\bibliographystyle{mnras}
\bibliography{draft} 

\appendix

\section{Additional figure}
\label{appendix}

\begin{figure*}      
\centering
\includegraphics[width=2.\columnwidth]{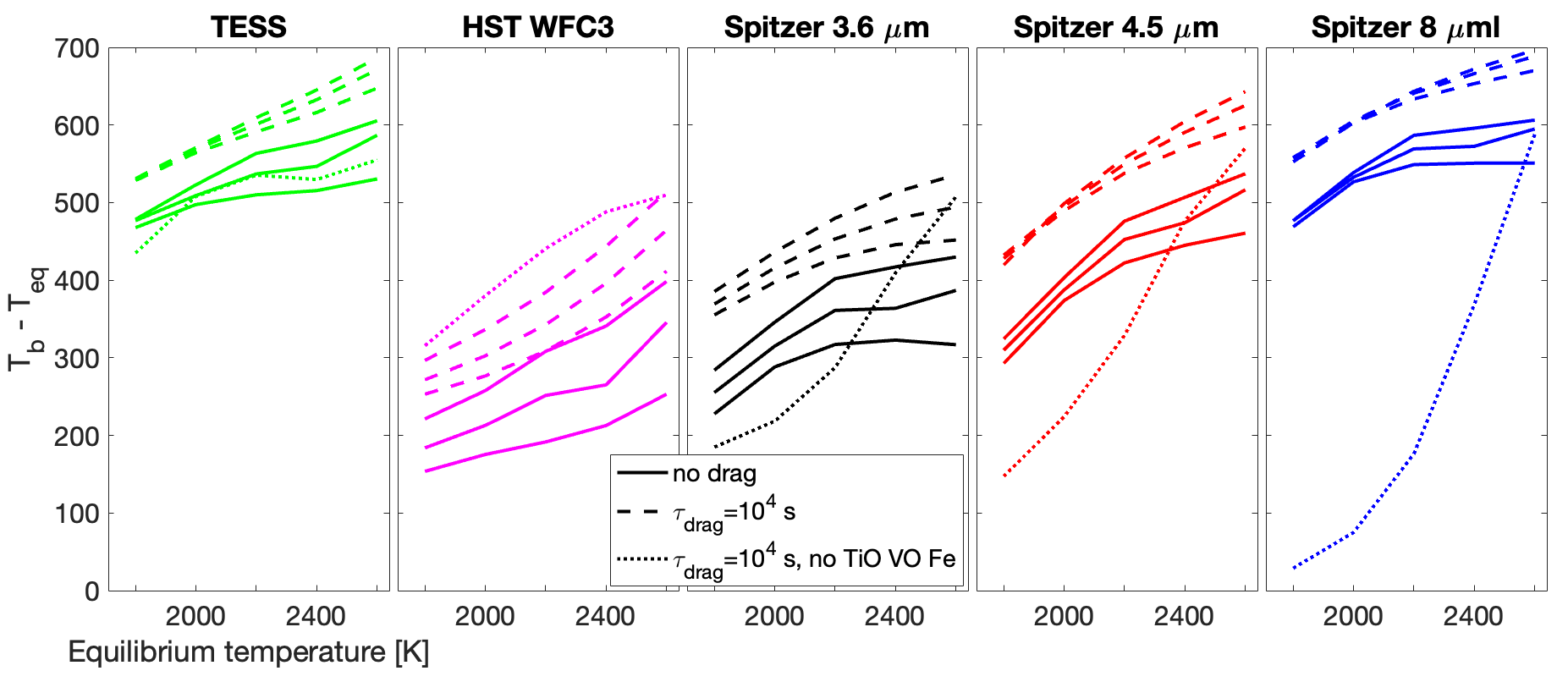}
\caption{Dayside brightness temperature from models as shown in Figure \ref{fig.compare_deming} but with additional outputs at the TESS band and without observational data for better visualization. Solid lines are models with no drag and three stellar temperatures; dashed lines are models with $\tdrag=10^4$ s and three stellar temperatures; finally, the dotted line is from models with $\tdrag=10^4$ s, a stellar temperature of 6500 K but excludes TiO, VO and Fe. }
\label{fig.dayside_model}
\end{figure*}

\section{Varying metallicity and C/O ratio: a case study for TOI 1431b}
\label{sec:metallicity}

\begin{figure*}      
\centering
\includegraphics[width=1.8\columnwidth]{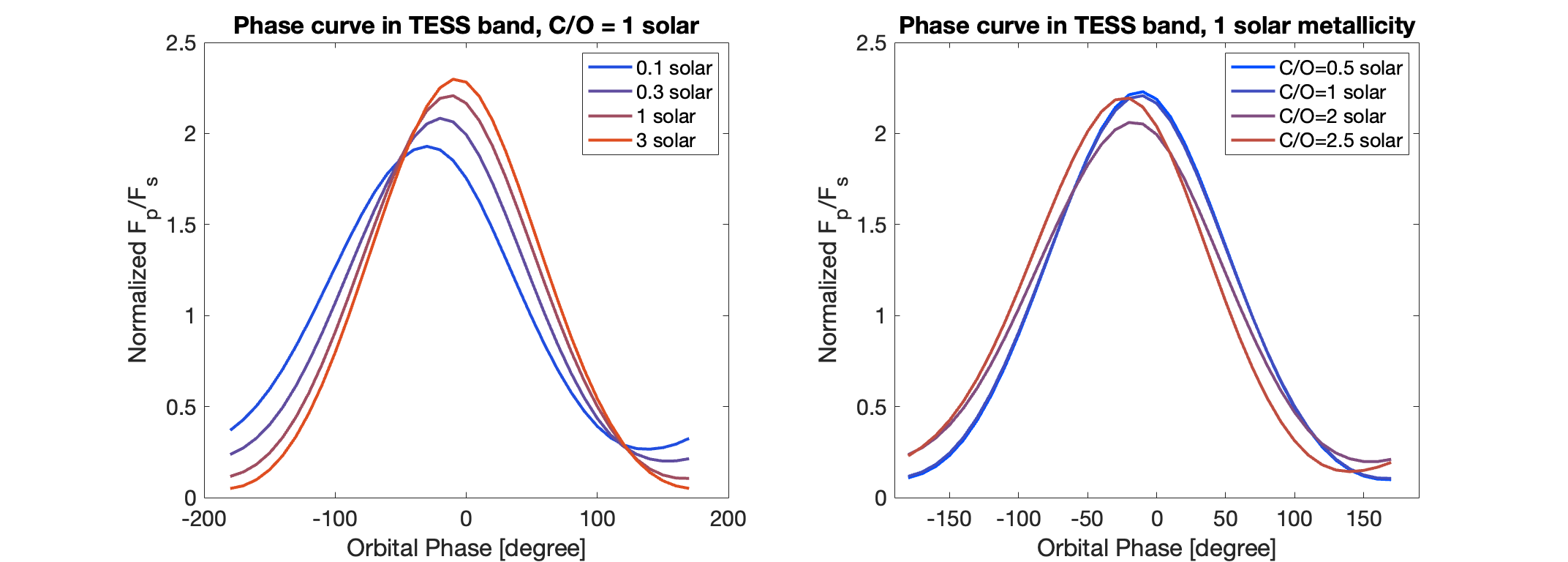}
\caption{Simulated phase curves in the TESS band from models appropriate for TOI 1431b. In the left panel, we show models with a fixed 1 solar C/O ratio but varying metallicity; in the right panel, we show those with a fixed 1 solar metallicity but varying C/O ratio.}
\label{fig.phasecurve_tess}
\end{figure*} 

\begin{figure}      
\centering
\includegraphics[width=0.9\columnwidth]{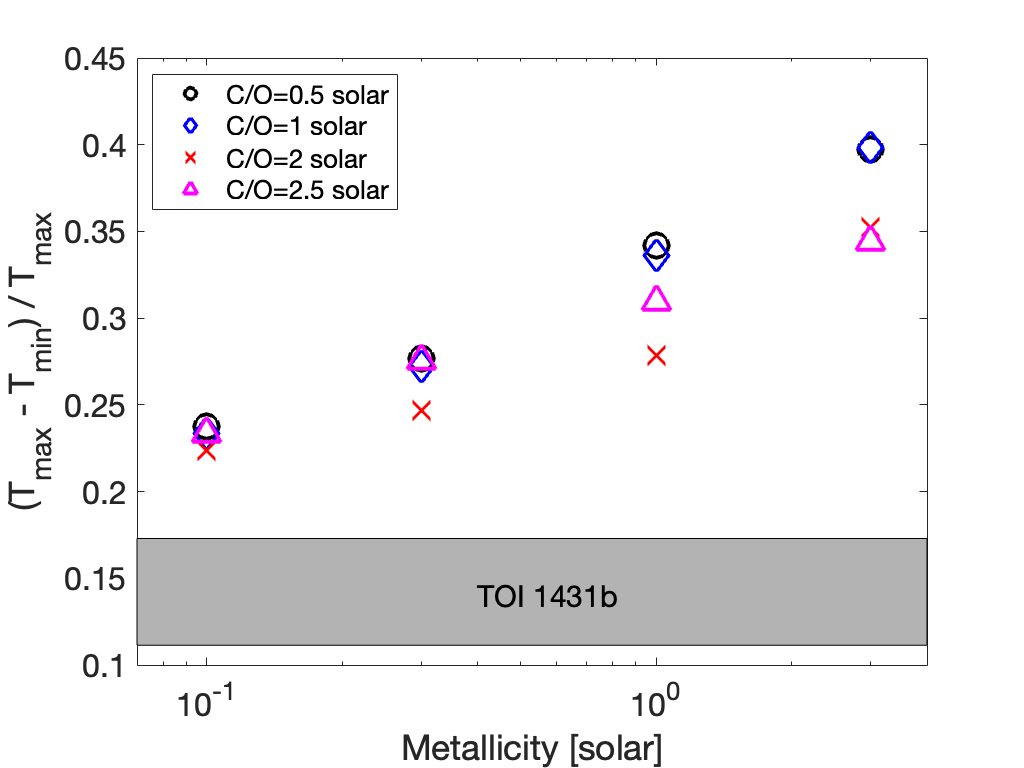}
\caption{The fractional day-night brightness temperature variations in the TESS band as a function of varying metallicity and C/O ratios for models of TOI 1431b.}
\label{fig.daynight_TOI1431b}
\end{figure}

\begin{figure*}      
\centering
\includegraphics[width=1.8\columnwidth]{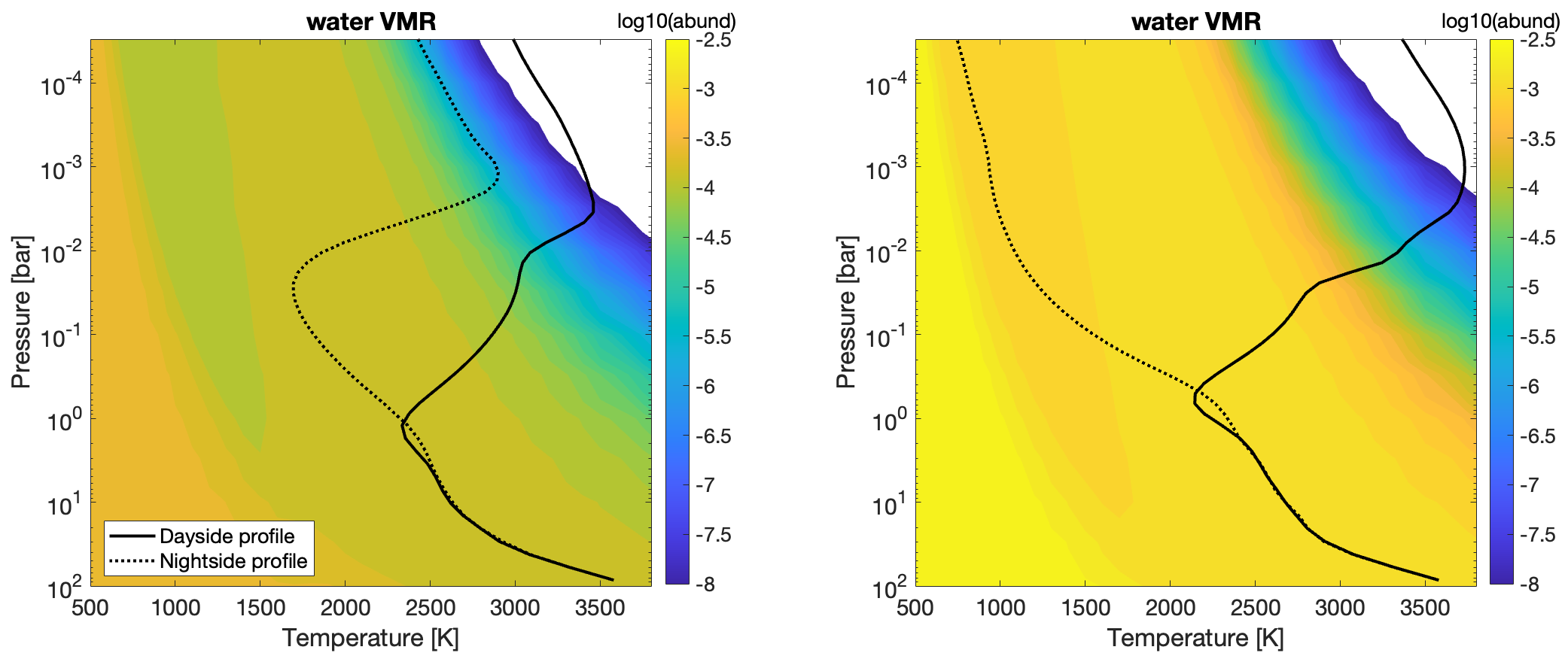}
\caption{Dayside and nightside TP profiles along with equilibrium water vapor volume mixing ratio in the temperature-pressure space for two models of TOI 1431b: on the left is the 0.3 solar metallicity case and on the right is the 3 solar metallicity case. Both cases have a 1 solar C/O ratio.}
\label{fig.water_TP}
\end{figure*} 

\begin{figure*}      
\centering
\includegraphics[width=2.\columnwidth]{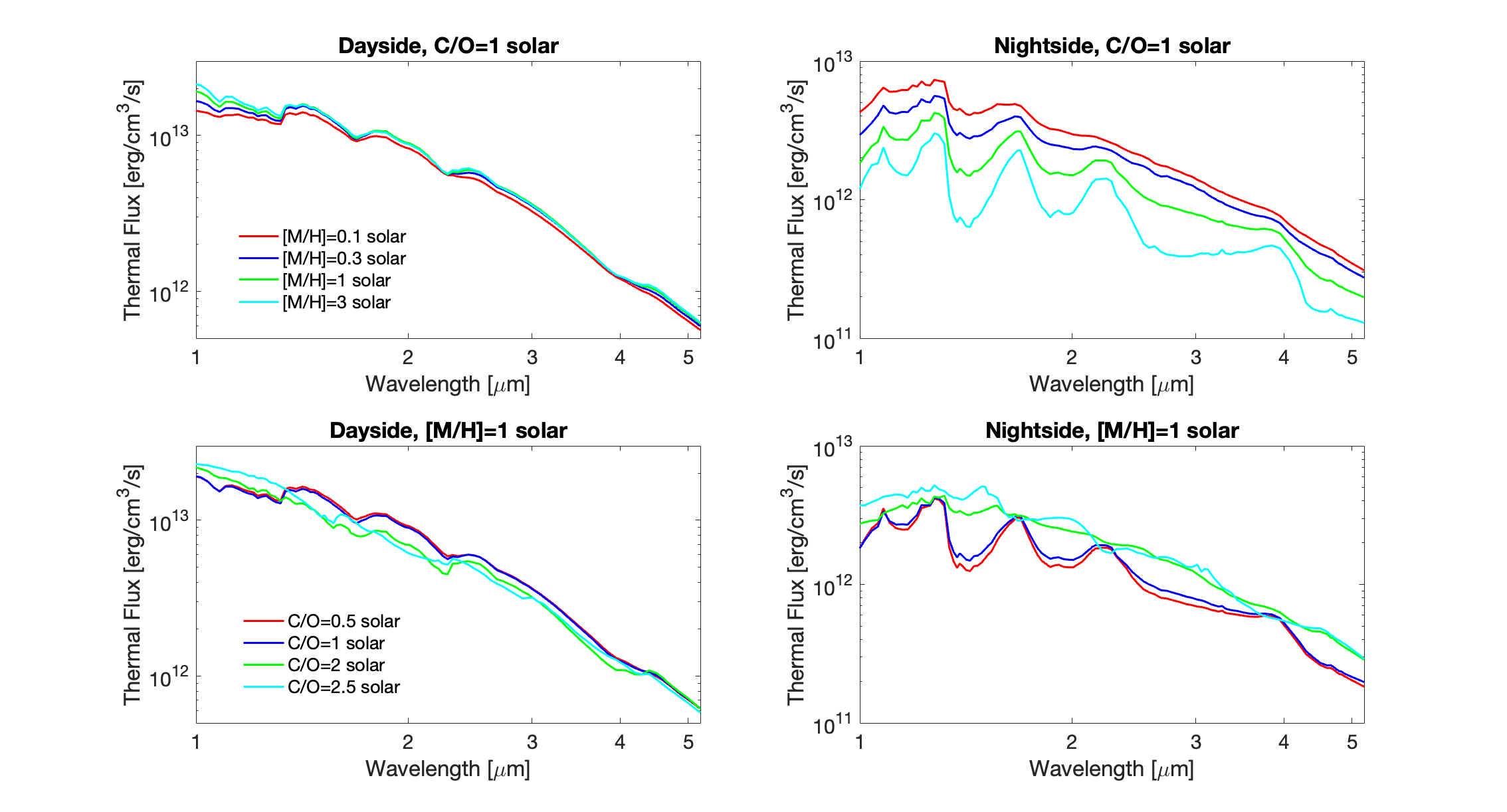}
\caption{Dayside (left column) and nightside (right column) emission spectra from models of TOI 1431b with varying metallicity and C/O ratio. Top row: models with a fixed 1 solar C/O ratio  but with varying   metallicity. Bottom row: models with a fixed 1 solar metallicity but with varying C/O ratio.}
\label{fig.spectra_toi1431b}
\end{figure*} 

The systematic overestimation of the modeled day-night temperature differences in the TESS band motivates us to look for alternative solutions to enhance heat transport. The majority of TESS targets that have day-night temperature contrasts much lower than model predictions do not have phase curve observations in other bandpasses, and so in this section, we  discuss TESS phase curve modeling only.  
We  vary the metallicity and C/O ratio to see if it would cause poor nightside radiative cooling that could lead to better day-night heat transport. Rather than modeling grids over planetary equilibrium and stellar temperatures which require significant computational resources, we focus on one planetary condition while changing the metallicity and C/O ratio. We use TOI 1431b as a template. Our previous work using a semi-grey GCM with radiatively active cloud tracers \citep{komacek2022} has also modeled this planet. 

We carry out a grid of models with  metallicity of 0.1 0.3, 1, and 3 solar, and C/O ratio of 0.5, 1, 2, and 2.5 solar. Note that in this set of GCMs we did not include opacities of Fe and a few other minor species in the k tables as was done in Section \ref{sec:results}. We have tried modeling this grid using the same k tables as in Section \ref{sec:results} but many of these models are computationally unstable. We then have to switch to a version of k tables without the Fe opacity that guarantee the stability of the GCMs. The main discrepancy is that there is no strong thermal inversion at very low pressure anymore which was supposed to be mainly driven by Fe absorption, and the stellar flux reaching the deeper IR photosphere is thus slightly more. Nevertheless, this is not expected to affect our qualitative conclusions.

We first show the resulting phase curves in the TESS bandpass from subsets of this grid in Figure \ref{fig.phasecurve_tess}. The normalized phase curves with a C/O ratio of 1 solar but varying metallicity are shown on the left panel. Higher-metallicity cases show higher day-night flux contrast and smaller phase-curve offsets and this trend is quite robust. In the right panel, with 1 solar metallicity, varying the C/O ratio does not have a clear trend and the over effect is less than varying the metallicity while holding the C/O ratio fixed as shown on the left. The trend that high metallicity generally results in larger day-to-night flux contrast and more aligned peak with the secondary eclipse is in agreement with previous studies of \cite{DobbsDixon2008,kataria2014,kataria2015}.

The normalized day-night temperature variations in the TESS bandpass of this grid are summarized in Figure \ref{fig.daynight_TOI1431b}, in which a grey shaded area represents the observed range of TOI 1431b. As seen in the phase curves, there is a general trend of increasing day-night temperature variations with increasing metallicity, while the varying C/O ratio adds on certain complexity. While low-metallicity GCMs significantly reduce the day-night temperature variation to about 0.23, this value is still significantly larger than the upper limit reported for TOI 1431b. A qualitatively similar trend of the day-night temperature variations also holds in other wavelengths which we do not further discuss here.

The trend could be understood via the competition between the day-night heat transport and radiative cooling/heating processes. The former is on the zeroth order held invariant for a given setup because the dominant heat transport near the photosphere is the hydrogen dissociation and recombination which is a priori independent of metallicity.  Radiative heating/cooling efficiency is our primary focus. Decreasing the metallicity generally results in a deeper photosphere simply because the opacity is positively correlated to the metallicity; therefore the longer radiative timescale  at higher pressures \citep{showman2002} naturally leads to smaller day-night temperature variations \citep{perezbecker2013,komacek2016}. Examples are shown in Figure \ref{fig.water_TP} in which 0.3 solar and 3 solar metallicities are displayed. The nightside structure at relatively low pressures is also significantly heated in the low-metallicity case (see the left panel of Figure \ref{fig.water_TP}). Without sufficient cooling agents in low-metallicity cases, for example, water content as shown in the color contours of Figure \ref{fig.water_TP} which is one of the most important coolants in giant planet atmospheres, the upper atmosphere is driven toward a strong thermal inversion. The hotter it gets, the less water to cool due to thermal dissociation, and this is positive feedback to drive a strong thermal inversion on the nightside of ultra-hot Jupiters. The other factor enhancing the nightside thermal inversion is the increasing heat transport efficiency by hydrogen dissociation and  recombination because of increasing atomic hydrogen mixing ratio with decreasing pressure. Whereas in the high-metallicity case on the right panel, water vapor is rich to sufficiently cool on the nightside and we see a decreasing temperature with decreasing pressure of the nightside profile that is typical in other hot Jupiter GCMs. 

Meanwhile, we are also cautious about the systematically low day-night temperature variations observed in the TESS band. Some studies on the same objects  claimed only the upper limits of the nightside flux which are smaller than those quoted in Figure \ref{fig.daynighttemp} (e.g., \citealp{wong2020systematic, wong2022variability}), indicating that the measurements could also be sensitive to data reduction methods. The significant mismatch of the GCM-derived day-night variations to the observed TESS values requires a better understanding of the data reduction processes in addition to searching for more theoretical possibilities. Nonetheless, this issue motivated the exploration of the strong dependence of hot Jupiter circulation on atmospheric compositions. 

Lastly, we show the prediction of the dayside and nightside emission spectra obtained from a subset of the models with varying metallicity and the C/O ratio in Figure \ref{fig.spectra_toi1431b}. With a 1-solar C/O ratio, varying metallicity in the top panels merely changes the size of the emission features on the dayside and the absorption features on the nightside. With a 1-solar metallicity, cases with 0.5 and 1 solar C/O ratios in the bottom panels show emission on the dayside and absorption on the nightside. However, cases with 2 and 2.5 solar C/O ratios are qualitatively different, lacking features in the water band, and are significantly closer to the blackbody spectrum. This is presumably because of the lack of sufficient water vapor in the atmosphere, and other opacity sources such as that of H$^-$ contribute more to the near-IR wavelength. The super solar C/O ratio to explain the much more muted features has been proposed for the HST dayside spectrum and {\it Spitzer} photometry  of WASP-18b \citep{sheppard2017} based on 1D retrievals for only the dayside (although recently been ruled out by JWST observations \citealp{brightstar2023}).  The spectral behaviors for super solar C/O ratio appear also on the nightside which is partly driven by the atmospheric heat transport and its effect on the nightside thermal inversion, and this should provide a unique test case for testing atmospheric heat transport with varying C/O ratio and metallicity.

\bsp	
\label{lastpage}
\end{document}